**Data-driven quantification of robustness and sensitivity of cell signaling networks**

Sayak Mukherjee[1,2], Sang-Cheol Seok[1], Veronica J. Vieland[1,2,4], and Jayajit Das[1,2,3,5*]

[1]Battelle Center for Mathematical Medicine, The Research Institute at the Nationwide Children's Hospital and Departments of [2]Pediatrics, [3]Physics, [4]Statistics, [5]Biophysics Graduate Program, The Ohio State University, 700 Children's Drive, Columbus, OH 43205.

**Abstract:**

Robustness and sensitivity of responses generated by cell signaling networks has been associated with survival and evolvability of organisms. However, existing methods analyzing robustness and sensitivity of signaling networks ignore the experimentally observed cell-to-cell variations of protein abundances and cell functions or contain ad hoc assumptions. We propose and apply a data driven Maximum Entropy (MaxEnt) based method to quantify robustness and sensitivity of *Escherichia coli* (*E. coli*) chemotaxis signaling network. Our analysis correctly rank orders different models of *E. coli* chemotaxis based on their robustness and suggests that parameters regulating cell signaling are evolutionary selected to vary in individual cells according to their abilities to perturb cell functions. Furthermore, predictions from our approach regarding distribution of protein abundances and properties of chemotactic responses in individual cells based on cell population averaged data are in excellent agreement with their experimental counterparts. Our approach is general and can be used to evaluate robustness as well as generate predictions of single cell properties based on population averaged experimental data in a wide range of cell signaling systems.

## Introduction

Robustness of specific cell functions against intra- and extra- cellular perturbations is a salient feature of many biological systems[1-4]. For example, polarization of selected molecules across wide variations of protein abundances in yeast cells is necessary for mating and bud formation[2]; similarly, the ability of *E. coli* cells to migrate to a nutrient rich region over a large range of nutrient concentration is related to their increased growth[4, 5]. Mutations increasing robustness of tumor cell survival and proliferation underscores the importance of robustness in diseases such as cancer[6]. Robustness in biological systems is often accompanied with processes that respond sensitively to perturbations[7, 8]. A striking example of this "robust yet fragile" behavior is found in individual T cells, which can accommodate over tenfold variations in protein abundances[9], yet also mount binary all or none responses to pathogenic ligands based on the formation of non-covalent T cell receptor ligand complexes with lifetimes differing from each other by just few seconds[10-12]. These examples suggest that both robustness and sensitivity (lack of robustness) of specific cell functions are key to understanding the design principles underlying cell signaling and gene regulatory networks.

Measuring robustness, however, raises a number of challenges. First and foremost is the inability of standard methods to incorporate experimentally observed cell-to-cell variations of protein abundances and cell responses[5, 13, 14]. For example, ordinary differential equation (ODE) based models, used to describe deterministic signaling kinetics of concentrations of signaling molecules averaged over a cell population, ignore intrinsic stochastic fluctuations[13, 15, 16] in individual cells that occur due to thermal fluctuations in biochemical signaling reactions[1, 4, 17, 18]. Similarly, these models generally do not account for extrinsic noise fluctuations[16, 19] arising from cell-to-cell variations of steady state abundances of signaling proteins and the physical properties of the signaling environment (e.g., cell size, molecular crowding, number of spatial compartments, or spatial micro-domains of receptors). But the unaccounted for variation of copy numbers of the signaling molecules produced by these fluctuations can generate qualitatively different signaling outcomes in single cells compared to those predicted by the ODE models for a cell population[20-22]. Thus these models might not properly reflect the behavior of the signaling network in single cells or even the behavior of a cell population. Methods have been proposed for including intrinsic and extrinsic noise fluctuations in stochastic and spatially resolved in silico models [2]. However, these methods still do not incorporate cell-to-cell variations of cell responses and model selection requires imposition of ad hoc criteria. For example, in quantifying robustness of different models designed to produce spatial polarization of marker molecules, the models that produced a polarity score of greater than an arbitrary threshold value of 0.6 were considered to be able to produce polarization, whereas, the corresponding experiments with synthetic circuits showed a wide distribution of the polarity score ranging between 0 and values larger than 1[2]. This can lead to erroneous conclusions regarding relative robustness of competing models, especially when the models show similar behavior. There are in addition computational challenges, whether working with deterministic or stochastic models, because of the large size of the parameter space, with many parameters required to describe strengths of interactions, total protein concentrations, and/or diffusion constants. Recent studies have proposed combining local Monte Carlo moves with principal component analysis (PCA)[18] or random walks in the parameter space[17] to address this issue.

Here we propose a novel data-driven approach based on maximum entropy (MaxEnt), a technique championed by Jaynes[23, 24], and maximum caliber (MaxCal) [25] to addresses these challenges. Our approach is entirely guided by available experimental data, measured in either a cell population or in individual cells, and it naturally combines intrinsic and extrinsic noise fluctuations in the cell signaling kinetics. MaxEnt has been widely used in diverse disciplines ranging from physics [26] to information theory [27] to biology [28-32] to estimate probability distributions of variables subject to constraints [24, 28, 30, 31]. In essence, the MaxEnt procedure yields the least structured, or least restricted, probability distribution for the underlying parameters, consistent with constraints imposed by available empirical data (such as average values). MaxCal [25, 33], also proposed by Jaynes, is an extension of MaxEnt to dynamical systems. These methods simultaneously allow us to directly incorporate stochastic properties of real networks, to avoid imposition of additional ad hoc modeling assumptions, and to bypass computational difficulties for some competing approaches. They also generate predictions regarding the distribution of specific attributes (e.g., adaptation time) of signaling kinetics in single cells, based on experimental observations that are only available at the cell population level. This addresses a common problem in inferring mechanisms underlying signaling kinetics in single cells. But beyond simply addressing technical problems, we also propose that MaxEnt and MaxCal uniquely provide a direct measure of biological robustness. The existing methods analyze sensitivity and robustness of specific cell functions as model parameters are perturbed[1, 2, 4, 18, 34]. In contrast, we constrain cell population averaged measurements in our MaxEnt approach to infer how model parameters are varying in individual cells, and to test predictions using the inferred probability distribution with available single cell measurements. Therefore, a consistency between our approach and the existing methods will suggest evolutionary selection of model parameters depending on how their perturbations influence model robustness. If this is correct for a biological system, then when applied to compare multiple mechanistic models underlying robust systems, the most robust MaxEnt or MaxCal model should coincide with the correct model. The ability of our approach to predict distributions of single cell attributes in a cell population allows us to test against single cell experiments if the most robust model is indeed the correct model. In order both to illustrate the approach and also to establish a proof of principle, we apply our method to quantification of the robustness of *Escherichia coli* (*E. coli*) chemotaxis.

The chemotactic behavior of *E. coli* cells is one of the best-characterized models of cell signaling to date[35, 36]. *E. coli* cells sense the presence of attractants (or repellants) in the medium with the help of transmembrane Tar receptors, and respond by swimming towards (or away) from the nutrient source. Upon binding to attractants, Tar receptors initiate a series of signaling events that lead to an increased bias towards anti-clockwise rotations of the flagella motor causing directed movements in individual *E. coli* cells[35, 36]. Then as the bacterial cells arrive at the region of higher nutrient concentrations, the flagella motor movements are biased towards clock-wise directions, and the cells return to their pre-stimulus state of random movements. This represents a perfect adaptive behavior, which has been demonstrated to be robust against variations of signaling protein concentrations, nutrient concentrations, rate constants, temperature changes, and strengths of interactions between proteins, both in experiments and in silico modeling [1, 5, 37, 38]. Recent experiments also provide data regarding cell-to-cell variations in chemotactic responses and protein abundances [5, 14, 39], in addition to the vast amount of cell population-averaged measurements carried out over many decades [40]. Therefore, *E. coli* chemotaxis is an ideal system in which to test and validate our approach.

We consider three different coarse grained or approximate models that have been proposed to explain E. coli chemotaxis: (i) the fine-tuned (FT) model [41, 42], (ii) the Barkai-Leibler (BL) model [1], and (iii) the modified BL (MBL) model, which is a recently proposed modification of the BL model [5]. Both BL and MBL can describe the robustness of the nearly perfect adaptation behavior in *E. coli* to variations of the interaction strengths and nutrient concentrations. However, BL fails, whereas MBL succeeds in describing how the cells manage to restrain pre-stimulus steady state values of the CheY-P protein within the working range of the flagella motors while being subject to wide variations in the abundances of the chemotactic proteins. The FT model fails to capture the robustness of *E. coli* chemotaxis and can reproduce the adaptive behavior only over a small range of parameters. We use stochastic simulations of the chemotaxis signaling network in individual *E. coli* cells, including intrinsic and extrinsic noise fluctuations, and use MaxEnt and MaxCal to quantify robustness of each of the three models, utilizing available cell population averaged and single cell data from the published literature as constraints. We also compare the models' predictions regarding distributions of chemotactic responses and protein abundances in individual cells with available experimental data.

The remainder of the paper is organized as follows. In section I we describe our overall approach and study design. Section II presents results regarding the relative robustness of the three models. Section III compares predictions of distributions of single cell adaptation time and protein concentrations with results obtained from experiments. Section IV is reserved for the discussion, while detailed methods are found in Materials and Methods and supplementary material.

**I. Approach and Study Design:**

Here we describe our approach in the context of chemotactic responses in individual *E. coli* cells responding to attractants added to the medium. The method can be generalized to any other cell signaling system where signaling kinetics data are available at the cell population and/or the single cell level.

*E. coli chemotactic models*: We consider three different coarse grained or approximate signaling models that were proposed to explain *E. coli* chemotaxis. Each model is composed of a set of biochemical reactions involving the chemotactic proteins, CheA, CheR, CheB, CheY, CheZ (only for MBL and FT), and the transmembrane Tar receptors. The models also differ from each other due to differences in molecular wiring between chemotactic proteins (Fig. 1). The kinetics of the chemotactic response for a given model is determined by the rate constants of different biochemical reactions, and also the total abundances of the associated proteins. The FT model (Fig. 1A) was among the first proposed models that could explain precise adaptation in *E. coli* within a narrow range of rate constants and protein concentrations [41, 42]. The BL model (Fig. 1B), proposed later, was able to capture the robustness of the nearly perfect nature of adaptation of *E. coli* chemotaxis to variations in rate constants or strengths of protein-protein interactions [1], as confirmed in cell population experiments [37]. However, the steady state concentration of phosphorylated CheY protein (or CheY-P) in the BL model is sensitive to large variations of protein concentrations in the model. Since the working range of the flagella motor is limited to small variations (~30%) from the optimal CheY-P concentration, the robustness of *E. coli* adaptation to large range of variation in the abundance of chemotactic proteins requires small variations of steady state CheY-P concentrations[5, 43]. Sourjik and colleagues proposed and experimentally tested a modified version of the BL model (MBL, Fig. 1C) that was able to restrain variations in CheY-P concentration to a small range [5].

All three models include biochemical reactions that describe the basic signaling events underlying *E. coli* chemotaxis (Fig. 1). Transmembrane Tar receptors bind to attractants (or repellants) in the medium and become de-active (or active). The kinase CheA, associated with activated Tar receptors, gets auto-phosphorylated and then transfers the phosphoryl group ($PO_4$) to the cytosolic enzyme CheY rendering the latter active. Phosphorylated CheY controls rotational bias of the flagella motor, an increase in CheY-P concentrations leads to an enhanced clockwise bias of the flagella motors [43, 44], causing individual *E. coli* cells to tumble. Tar receptors also undergo methylation by an enzyme, CheR, and the enzyme CheB removes the methyl groups from the active form of the receptors. Methylated receptors induce CheY phosphorylation at an increased rate.

When attractants are added to the medium, a decrease in the number of activated receptors leads to decrease in CheY phosphorylation, and the *E. coli* cells display a chemotactic response by executing a directed motion toward the nutrient source. However, the decrease in active receptor number results in an increase in the rate of methylation of the receptors due to the decreased rate of de-methylation by CheB. As methylation goes up, the activity of the receptors and the phosphorylation of CheY start to increase and eventually return to the pre-stimulus level, resulting in the restoration of random motion in *E. coli*. CheY-P concentrations return almost precisely to the pre-stimulus level after a time interval, which generates the nearly perfect adaptive behavior.

The key difference in signaling between the FT and the BL model is that the enzyme CheB can demethylate only the active methylated receptors in BL, whereas, CheB demethylates both the active and the inactive receptors in the FT model (Fig. 1). This induces an integral feedback control process in the BL model[45], where the production of active methylated receptors, the key inducer of CheY-P, is regulated by the sum of the difference between the actual abundance of active methylated receptors with its steady state value over a time period. This produces a steady state of active methylated receptors independent of the ligand concentrations for a wide range of model parameters in the BL model. In the absence of such a control mechanism, the FT model displays perfect adaptation only in a small range of parameters. However, since CheY-P undergoes auto dephosphorylation (Fig. 1) in the BL model, increases in the total abundances of enzymes such as CheA also lead to an increase the steady state concentrations of CheY-P. In contrast, CheY-P is de-activated by the enzyme CheZ in the MBL model (Fig. 1), thus, any change in the rate of CheY-P production due to changes in abundances of phosphorylating enzymes such as CheA can be counteracted by a correlated change in abundances in phosphatases such as CheZ[5]. This produces an increased robustness of steady state abundances of CheY-P in the MBL model. The specific biochemical reactions for each model are described in Fig. 1 and the supplementary material.

*Quantification of robustness of E. coli chemotaxis using Maximum Entropy:* We consider signaling kinetics in individual *E. coli* cells, where copy numbers of signaling proteins can vary due to intrinsic noise fluctuations. In addition, total protein abundances vary from cell to cell, representing the effects of extrinsic noise fluctuations. We assume that the rate constants are primarily affected by the thermodynamics of the protein-protein interactions and do not vary from cell to cell. Here we restrict our attention to spatially well mixed models. However, the methods can easily be extended to include spatial events (such as receptor clustering) that can affect signaling kinetics in *E. coli* chemotaxis [46] as well as in other signaling systems.

Upon addition of attractants in the medium at time $t = t_0$, in an individual cell containing total protein abundances given by $\{n^{total}_q\}$ (q=1… $N_T$, representing the proteins Tar, CheA, CheB, CheR, CheZ, and CheY), the copy numbers of signaling molecules change with time due to the chemotaxis signaling reactions. We define a stochastic trajectory, Γ, representing kinetics of the abundances of signaling proteins in an individual cell in a time interval $t_0$ to $t_n$ by a set ($\{n_j\}$, $t_n$ ; $\{n_j\}$, $t_{n-1}$ ; $\{n_j\}$, $t_{n-2}$; ….; $\{n_j\}$, $t_1$ ; $\{n_j\}$, $t_0$ ; $\{n^{total}_q\}$). $\{\{n_j\},t_{n-i}\}$ denotes copy numbers of $N_P$ different proteins (j=1…$N_P$) at any time $t_{n-i}$ (= $t_0$+(n-i)Δ, i=0..n) ; $\{n_j\}$ includes modified protein species generated during signaling such as CheY-P, as well as, unmodified protein species such as CheY. The time interval Δ is chosen to be smaller than or of the same order of the smallest reaction time scale (Fig. 2). We have $N_P \geq N_T$, as a protein species can be modified during signaling, e.g., the signaling protein CheY-P is generated from the protein CheY. The properties of chemotactic responses in a single cell, such as the adaptation time or the precision of adaptation, depend on the stochastic trajectory Γ of that cell (Fig. 2A). Therefore, we represent the magnitude ($f_a$) of a specific property, $a$ (e.g., adaptation time), of the chemotactic response displayed by a single cell by $f_{aΓ}$. Denoting the probability that a single cell follows a stochastic trajectory, Γ, be $P_Γ$, the distribution of $f_a$ can be written as

$$p(f_a) = \sum_{\{Γ\}} \delta_{f_a, f_{aΓ}} P_Γ \qquad (1)$$

where the summation is over a set of stochastic trajectories {Γ} observed in a population of *E. coli* cells. $P_Γ$ can be calculated from the joint distribution P($\{n_j\}$, $t_0$ ; $\{n^{total}_q\}$) of the signaling proteins at the pre-stimulus level and the total protein abundances first, and then by calculating the transition probabilites with which the initial state ($\{n_j\}$, $t_0$) goes to the consecutive signaling states at different times. Therefore we have

$$P_Γ = P(\{n_j\}, t_n ; \{n_j\}, t_{n-1} ; …. ; \{n_j\}, t_1 | \{n_j\}, t_0) P(\{n_j\}, t_0 ; \{n^{total}_q\}) \qquad (2)$$

The biochemical reactions in each model occur as Markov processes where the probability P($\{n_j(t_2)\}|\{n_j(t_1)\}$) ($t_2 \geq t_1$), that the state changes from ($\{n_j\}$, $t_1$) to ($\{n_j\}$, $t_2$) is given by the Master Equation[15],

$$\frac{\partial P(\{n_j(t_2)\}|\{n_j(t_1)\})}{\partial t_2} = LP(\{n_j(t_2)\}|\{n_j(t_1)\}) \qquad (3)$$

In Eq. (3), $L$ is a linear operator [15] dependent on the biochemical reaction rates, wiring of the signaling network, and the numbers of signaling proteins at time $t_1$.

The steady state conditional probability of the abundance of signaling molecules at the pre-stimulus state at $t = t_0$ given a fixed set of total protein abundances or P($\{n_j\}$, $t_0$ | $\{n^{total}_q\}$) can be calculated from Eq. (3) by setting the left hand side to zero and using the

form for *L* without any attractant present in the medium. The joint probability distribution, $P(\{n_j\}, t_0; \{n^{total}_q\})$ then can be calculated using the relation $P(\{n_j\}, t_0; \{n^{total}_q\}) = P(\{n_j\}, t_0 | \{n^{total}_q\}) P(\{n^{total}_q\})$, where, $P(\{n^{total}_q\})$ denotes the distribution of total protein abundances in single cells.

When experimental measurements are available for a population of cells, e.g. from western blot assays, data are available as average values of the properties of the chemotactic response,

$$\frac{1}{\text{total \# of cells}} \sum_{\alpha=1}^{\text{total \# of cells}} f_a^\alpha = \sum_{\{\Gamma\}} f_{a\Gamma} P_\Gamma = \overline{f_a^{expt}} \qquad (4)$$

where $\overline{f_a^{expt}}$ denote the experimental average value for the $a^{th}$ property of the chemotactic response. Note that $\sum_{\{\Gamma\}} f_{a\Gamma} P_\Gamma$ refers to the expected value of the variable, $f_a$, rather than the sample average. Therefore, the equality in Eq. (4) is strictly true only as the number of samples (i.e., of cells) becomes very large[47]. In finite samples, the average can deviate from the expected value, potentially leading to errors in the estimated distributions of the parameters (see the Discussion section for further details). Note too that to make our notation clear, as previously noted, we will use $f_a$ to denote properties of chemotactic responses (e.g., adaptation time or precision of adaptation), and $n^{total}_q$ for protein abundances in individual cells. Single cell measurements can provide further details regarding how a specific property is distributed in a cell population, and they could allow for the calculation of variances or even higher order moments from the experimental data, i.e.

$$\frac{1}{\text{total \# of cells}} \sum_{\alpha=1}^{\text{total \# of cells}} (f_a^\alpha)^n = \sum_{\{\Gamma\}} f_{a\Gamma}^n P_\Gamma = \overline{\left(f_a^{expt}\right)} \qquad (5)$$

where the right hand side denotes average of the *n*th moment for the property $f_a$ calculated from experimental data. In addition, average values of the total protein abundances from cell population level experiments, variance, and higher moments for the total protein numbers from single cell measurements, might be available, i.e.,

$$\frac{1}{\text{total \# of cells}} \sum_{\alpha=1}^{\text{total \# of cells}} (n_q^{total})_\alpha = \sum_{\{n_q^{total}\}} n_q^{total} P(\{n_q^{total}\}) = \overline{n_q^{expt}}$$

$$\frac{1}{\text{total \# of cells}} \sum_{\alpha=1}^{\text{total \# of cells}} (n_q^{total})_\alpha^n = \sum_{\{n_q^{total}\}} (n_q^{total})^n P(\{n_q^{total}\}) = \overline{(n_q^{expt})^n}$$

$$(6)$$

where $n_q^{total}$ denotes the total abundance of protein $q$ in a single cell. The probability function $P_\Gamma$ contains variations due to intrinsic and extrinsic noise fluctuations through $P(\{n_j\}, t_n ; \{n_j\}, t_{n-1} ; \ldots ; \{n_j\}, t_1, \{n_j\}, t_0 | \{n^{total}_q\})$ and $P(\{n^{total}_q\})$. Therefore, it is possible to choose different shapes of distributions of the total number of proteins or $P(\{n^{total}_q\})$, which will satisfy the constraints imposed by the properties of the chemotactic response and/or the available cell population and the single cell data for total protein abundances. We seek to estimate the maximally varying, or the least structured distribution of the stochastic signaling trajectories, where the minimal structure in the distribution of the protein abundances arise solely due to the constraints imposed by the available experimental data. Such a distribution represents the maximal cell-to-cell variations the system can endure while reproducing the experimentally measured data at the single cell and the cell population level.

In order to estimate this distribution which incorporates the available experimental data, we maximize the Shannon entropy ($S$) constructed from $P_\Gamma$ [33]

$$S = -\sum_{\{\Gamma\}} P_\Gamma \ln P_\Gamma \qquad (7)$$

in the presence of the constraints imposed by Eqn. (4)-(6). We refer to the resulting distribution, $\widehat{P_\Gamma}$, as the constrained MaxEnt distribution. We reiterate that while searching for a MaxEnt solution we considered different $P_\Gamma$ arising from different distributions of the total protein concentrations and selected the $P_\Gamma$ that satisfied all the imposed constraints and produced the maximum value of $S$. More details on the implementation can be found in the methods section and the supplementary material. In order to compare distribution of protein total abundances, $\hat{P}(\{n_q^{total}\})$ corresponding to $\widehat{P_\Gamma}$ with the unconstrained case, we construct a uniform distribution of the total protein concentration $\hat{Q}(\{n_q^{total}\})$. We chose the uniform distribution as it has the maximum uncertainty. Maximization of the path entropy, defined in Eqn (7), is also known as the maximum caliber distribution, which we refer to as the MaxCal [25, 33], which can also be derived in either the constrained or the unconstrained form.

Because the unconstrained MaxEnt (or MaxCal) represents the greatest robustness available to a given model in the absence of any data, while the constrained MaxEnt (MaxCal) represents the greatest robustness available to the model while constraining its behavior to conform to experimental results, the *difference* between the two is a measure of the degree to which the model must deviate from the uniform distribution in order to accommodate the data, or in other words, how great a departure from the uniform distribution is required to bring the model into accordance with empirical observations on the behavior of the system. Thus the model exhibiting the maximum robustness is the model with the minimal relative entropy (MinRE),

$$\text{MinRE} = \sum_{\{n_q^{\text{total}}\}} \hat{P}(\{n_q^{\text{total}}\}) \ln[\hat{P}(\{n_q^{\text{total}}\})/\hat{Q}(\{n_q^{\text{total}}\})] \tag{8}$$

Note that MinRE is a particular form of the Kullback-Leibler distance[48] in which the probability ratio reflects the constrained versus unconstrained MaxEnt distributions and the expected value of the ln probability ratio is taken with respect to the former. One advantage of using MinRE rather than Shannon's Entropy S (Eq. 7) in quantifying robustness is that, unlike S, the lower bound for MinRE is always zero, which makes MinRE a better metric than S.

Under the constraints in Eq. (4)-(6), the maximization of S will lead to an estimate of $\hat{P}(\{n_q^{\text{total}}\})$,

$$\hat{P}(\{n_q^{\text{total}}\}) = Z^{-1} \exp\left[-\sum_{a=1}^{r} \lambda_a \left(\sum_{\{\Gamma_C\}} f_{a\Gamma} P_C\right) - \sum_{a=1}^{m} \kappa_a \left(\sum_{\{\Gamma_C\}} (f_{a\Gamma})^n P_C\right)\right] \exp\left[-\sum_{q=1}^{N_1} \eta_q n_q^{\text{total}} - \sum_{q=1}^{N_2} \mu_q (n_q^{\text{total}})^n\right]$$
$$\times \exp\left[-\sum_{\{\Gamma_C\}} P_C \ln P_C\right] \tag{9}$$

, where, average values and the nth order moments of chemotactic properties, indexed as 1 to r and 1 to m, respectively, and, average values and the nth order moments of total abundances of proteins indexed by 1 to $N_1$ and 1 to $N_2$, respectively, have been constrained. In the above expression, $P_C = P(\{n_j\}, t_n ; \{n_j\}, t_{n-1} ; ....; \{n_j\}, t_0 \mid \{n^{\text{total}}_q\})$, is the conditional probability of generating a stochastic trajectory $\Gamma_C$ represented by $\{\{n_j\}, t_n ; \{n_j\}, t_{n-1} ; \{n_j\}, t_{n-2} ;....; \{n_j\}, t_1; \{n_j\}, t_0\}$ given a fixed set of total protein abundances, $\{n_q^{\text{total}}\}$. The sum over $\{\Gamma_C\}$ essentially denotes averages over variations of stochastic trajectories due to intrinsic noise fluctuations. The Lagrange's multipliers in the above equation, $\{\lambda_a\},\{\kappa_a\},\{\eta_q\}$, and $\{\mu_q\}$, are calculated by substituting the estimated $\hat{P}(\{n_q^{\text{total}}\})$ in Eq. (9) in the equations for the constraints, and then solving the resulting system of nonlinear equations for the Lagrange multipliers. The derivation of Eq. (9) is shown in the supplementary material.

The above distributions provide characterizations of the properties of chemotactic responses in individual cells. In addition, $\widehat{P_\Gamma}$ can be used to evaluate the distributions of total protein concentrations in a cell population. We compare these predictions with the data available in single cell experiments in section III. Additional details regarding implementation are shown in Methods and Materials and the supplementary material.

*Study Design*: In order to assess robustness of different models, we simulated stochastic biochemical signaling processes in the FT, BL, and the MBL models using a continuous time Monte Carlo method also known as the Gillespie method [49]. Variations of steady state protein abundances in individual cells due to extrinsic noise fluctuations were also considered. The total protein concentrations were chosen from uniform distributions in an interval [0 $U_H$], where the value of $U_H$ assumes different values for different proteins. The rate constants in the simulations were set to their measured and estimated values obtained from the literature [5]. Since the rate constants are primarily determined by the thermodynamics of protein-protein interactions, we did not consider variations of these constants in individual cells in the simulations. See Materials and Methods for further details. We considered two types of constraints imposed by the available experimental data. Type (A) constraints were imposed by the observed characteristics of the chemotactic response. We used three different properties of chemotactic response that have been measured in experiments: (1) adaptation time, $\tau$, defined as the time the abundance of CheY-P in an individual *E. coli* cell takes to rise up to half of its pre-stimulus value from the time when attractants were added; (2) precision of adaptation, *s*, calculated as the absolute value of the relative difference in the steady state abundances of CheY-P at the pre- and post-stimulation conditions; (3) variation of the pre-stimulus abundance of CheY-P in the steady state (or *p*) relative to its value at the optimal condition. Type (B) constraints involved constraining protein abundances using their average values in a cell population, or distributions available from single cell experiments. We used experimentally measured [40] values for average concentrations of the proteins used in the models, Tar, CheA, CheR, CheB, CheY, and CheZ, and cell population averaged chemotactic responses. In addition, we used data from single cell experiments measuring distributions of CheY, joint distributions of CheY-CheZ and CheY-CheA [5], and the distribution of the adaptation time in individual *E. coli* cells.

## II. Quantification of robustness of *E. coli* chemotaxis:

We first considered the robustness of the three models when the average value of precision of adaptation (or $\bar{s}$) over a cell population was constrained (Fig. 3A). As expected from the robustness analysis of the models reported in the literature, the BL and MBL models produced substantially smaller values of MinRE compared to that of the FT model. This increased robustness of the BL and MBL models is due to the presence of an integral feedback control [45]. The MBL model produced slightly higher values of MinRE compared to the BL model, especially at very small values of the average precision. The reason for this is that the speed of adaptation depends on the total abundance of CheB (for BL) and CheB-P (for MBL) (See Supp Material). The models fail to adapt properly when the copy numbers of the enzyme (total CheB for BL and CheB-P for MBL) demethylating the active receptors become very small. Since the copy number of the phosphorylated form, CheB-P can be much smaller than that of CheB, it is more likely for MBL to generate cases that do not adapt (see Supplementary Material and Supplementary Figure 1).

Next we calculated the values of MinRE for the three models when the cell population averaged value of $\tau$ ($\bar{\tau}$) or $p$ ($\bar{p}$) was constrained to the experimentally observed values (Figs 3B,C). Similar to Fig. 3A, the FT model was much less robust compared to the BL and MBL models when $\bar{\tau}$ was held fixed at the experimentally observed value of 245 s (Fig. 3B). The adaptation module of the FT model lacks an integral feed forward mechanism, and consequently performs rather weakly in the face of variations in the protein concentrations. Thus, only a very narrow range of protein abundances can generate perfect adaptation. The MBL model again produced slightly higher values of MinRE compared to the BL model, due to the greater number of cells that did not adapt well compared to BL model. This occurs for the same reason mentioned above, as poor adaptation also leads to poor precision of adaptation. All three models produced comparable values of MinRE, with MBL and BL producing the smallest and the largest values, respectively, when the average $p$ values were constrained (Fig. 3C). This occurred due to the following reason. Since restricting the steady state abundances of CheY-P requires correlated variations of protein abundances in all the models [5], a relatively smaller subset of cells that were randomly assigned correlated variation of protein abundances when the protein abundances were initially drawn from uniform distributions were able to produce values of $p$ closer to the experiments. This raised the values of MinRE in all the models compared to the cases when $\bar{\tau}$ or $\bar{s}$ were constrained. The MBL model, designed to produce smaller variations of $p$ for correlated variations of protein concentrations, produced the smallest values of MinRE. The FT model produced lower values of MinRE compared to the BL model, as the presence of the phosphatase CheZ in the FT model ensures that the steady state concentration of CheY-P stays relatively resilient to the variations in the protein abundances[5].

Next we simultaneously constrained $\bar{s}$, $\bar{\tau}$, and, $\bar{p}$ (Fig 3D). The MBL model displayed smaller values MinRE or higher robustness behavior compared to the BL and the FT model. This is because the MBL model is designed to produce smaller values of $\bar{p}$, therefore, it can accommodate more variation in protein abundances than the other models while reproducing the average values of $\bar{s}$, $\bar{\tau}$ and $\bar{p}$. The constraint imposed by $\bar{p}$ has a greater role in regulating the MinRE values in the three-constraints case, since holding $\bar{p}$ fixed within small range requires substantially restricted variations in protein abundances. This is reflected in differences in MinRE of one or two orders of magnitudes in Fig. 3C compared to Figs. 3A and 3B. The BL model produced lower values of MinRE compared to the FT model. However, MinRE values were comparable for the BL and FT models. The large separation in MinRE values produced by the FT and the BL models when $\bar{s}$ or $\bar{\tau}$ was constrained is reduced in Fig. 3D, since constraining $\bar{p}$ to a small range (<30%) in both the models requires substantial restrictions in the variations of protein abundances (Figs. 3C and Supplementary Figure 2D). Therefore, when all three variables, $\bar{s}$, $\bar{\tau}$ and $\bar{p}$ were constrained both the FT and the BL model produced similar values of MinRE.

Finally, we constrained the cell population averaged value of $\tau^2$ (or $\overline{\tau^2}$) in addition to the average values of $\bar{s}$, $\bar{\tau}$ and $\bar{p}$ (Fig. 3E). We checked whether the value of $\overline{\tau^2}$ is independent of $\bar{\tau}$ by comparing experimental data for $\tau$ with an exponential or a Poisson distribution; a Gaussian distribution appears to be a better fit to P($\tau$) (details in supplementary Table 4 in the supplementary material) suggesting that $\overline{\tau^2}$ does not depend on $\bar{\tau}$. In this case, the relative rank ordering between the three models remains unchanged from the three-constraints case shown in Fig. 3D. However, the separation between the FT and the BL models increases, since in the FT model the cases that displayed poor adaptation also produced much larger values of $\tau$ compared to the BL model. Therefore, the FT model required a greater restriction in the protein abundances to reproduce the experimentally observed value of $\overline{\tau^2}$ compared to the BL model. The relative rank ordering of the models (MBL>BL>FT) based on the MinRE values remain unchanged (Supplementary Figure. S6A) when variances of p and s were further constrained in addition to the above constraints ($\bar{s}$, $\bar{\tau}$, $\bar{p}$, and $\overline{\tau^2}$).

Thus overall, the MBL model was found to be consistently the most robust when the chemotactic responses are constrained. In the following section, therefore, we restrict attention to this particular model.

**III. Predictions of distributions of the individual cell attributes:** We compared the predictions from (Eq. 9) regarding distributions of properties of the chemotactic response and specific protein abundances in individual cells with experiments for the MBL model. The predicted distribution for $\tau$ when $\bar{s}$, $\bar{\tau}$ and $\bar{p}$ were constrained to their experimental counterparts showed a wider distribution for $\tau$ compared to that observed in experiments (Supplementary Figure 3A). When the variance of $\tau$ was constrained along with, $\bar{s}$, $\bar{\tau}$ and $\bar{p}$, the predicted distribution agreed reasonably well with the experiments (Fig. 4A). However, we found that constraining $\bar{\tau}$ and $\overline{\tau^2}$ alone to their experimental values also produced a reasonable agreement between the predicted distribution of $\tau$ and the experiment (Supplementary Figure 3B). This behavior could arise if $\tau$ is not substantially correlated with *s* and *p* in individual cells for the set of output constraints investigated. To test this conjecture, we calculated the Pearson correlation co-efficients[50] ($r_{\tau s}$, $r_{\tau p}$ and $r_{sp}$) between τ, *s* and *p*, respectively, under the joint distribution of τ, *s*, *p* and $\tau^2$ when the values of $\bar{s}$, $\bar{\tau}$, $\bar{p}$ and $\overline{\tau^2}$ were constrained to their experimental values. We found that $r_{\tau s}$ has the largest value ($r_{\tau s}$ = 0.0349 <<1, $r_{\tau p}$ = 0.0087 and $r_{sp}$ = 0.0254), implying that τ is not strongly correlated with *s* and *p* in individual cells, resulting in distributions of τ that are primarily regulated by τ and $\tau^2$ in single cells.

Next, we compared the predicted distributions for protein abundances in single cells when $\bar{s}$, $\bar{\tau}$, $\bar{p}$ and $\overline{\tau^2}$ were constrained to their experimental values from the single cell experiments reported in [5, 37, 39]. We calculated the first six moments from the predicted distribution of CheY abundance in single cells and compared them with their experimental counterparts [5]. Fig 4B shows that the predicted distribution produced much larger values for the moments to those observed in the experiments. When we further constrained the averages, variances and the covariances of CheY and CheZ to the experimentally observed values, the predicted moments for CheY abundance again showed excellent agreement with experiments (Fig 4C). The higher moments deviate slightly upwards from the y=x curve, which can be indicative of the fact that the actual CheY distribution is a log normal distribution with a longer tail [5]. The predictions for CheZ abundances also showed similar behavior to that of CheY when $\bar{s}$, $\bar{\tau}$, $\bar{p}$ and $\overline{\tau^2}$ were constrained alone or in combination with average values and variances of CheY abundances in individual cells (Fig 4D and 4E). We note that the uncertainties in the values of the higher moments of the variables can be largely due to the small size of the available data. Variances were estimated assuming normal distributions for those variables[51] and we tested our predictions within those variances (+/- 1 standard deviation). We also checked the independence of the variances and covariances of the abundances of CheY and CheZ from the mean values (supplementary Table 4). The availability of additional data would improve our ability to further test these predictions.

## IV. Discussion

We showed that a data driven MaxEnt based approach can successfully quantify robustness of signaling models of *E. coli* chemotaxis. The robustness is measured considering cell-to-cell variations of protein abundances and chemotactic responses in a population of *E. coli* cells that reproduce the experimentally observed chemotaxis in the wild type *E. coli* strain RP437. This approach is markedly different from a class of traditional experimental or in silico methods where parameters regulating the cell function(s) are perturbed one at a time or simultaneously to measure robustness of a signaling system. The robustness of a model in such perturbation studies is quantified by considering the range of variation in the model parameters that can be accommodated without changing the output responses. The larger the range of the perturbations that can be tolerated, the greater is the robustness. Our MaxEnt based approach considers the range of variation of the parameters in individual cells such that the cell population is able to reproduce the experimentally observed population averaged cell signaling responses. In this approach, the larger the range of the cell-to-cell variation of model parameters, the greater is the robustness. We show that the MaxEnt based quantification of robustness (Fig. 3 and Supplementary Figure 2) is in agreement with the estimation of robustness using traditional perturbation studies. The MBL model turned out to be the most robust model against cell-to-cell variations occurring from intrinsic and extrinsic noise fluctuations, followed by the BL model, while reproducing experimentally observed chemotactic responses in individual cells. The FT model was found to be substantially less robust than either the MBL or the BL model. These results are consistent with the results from the previous studies investigating sensitivity of *E. coli* chemotaxis against variations of protein concentrations and kinetic rates using *in silico* modeling and overexpression experiments[1, 5, 37]. The agreement of the rank ordering of these models based on our MinRE measure with the existing robustness analysis for the three models we analyzed validates our approach. Furthermore, it shows that the relative robustness of these models remains unchanged when cell-to-cell variations of chemotactic responses as well and protein abundances are included in the quantification of robustness.

The agreement between the robustness of different models from the MaxEnt analysis and the traditional perturbation experiments and simulations points to an interesting issue in biology. Did robust cell signaling systems evolve to produce larger cell-to-cell variations in parameters that can accommodate large perturbations without changing cell responses? The agreement between the two methods points us to an affirmative answer. We further probed this question by comparing the single cell distributions predicted from our MaxEnt analysis with available single cell measurements in *E. coli* chemotaxis. We found that the single cell distributions of protein abundances of CheY and CheZ could produce the experimentally observed distributions when the average values, variances, and co-variances were constrained to the experimental data. The rank ordering of the robust models was consistent with that obtained from perturbation experiments when the above constraints were imposed. This provides further indication that evolutionary selection of abundances or magnitudes of parameters in a signaling network in individual cells is influenced by their ability to control the robustness and sensitivity of cell functions.

Our MaxEnt based approach is similar in spirit to some recent work in parameter estimation techniques for biochemical networks using Bayesian methods[52-54]. These methods infer distributions of parameters (e.g., rate constants in signaling networks) by evaluating the posterior distribution of the parameters given the available experimental data using Bayes' rule; models producing larger Bayes factor are then considered to be more appropriate for explaining the measured data[53, 54]. These methods require assumption of specific distributional forms for both priors and likelihoods. By contrast, the MaxEnt approach is free from such assumptions, so that the inference is solely guided by the available experimental data. The connection between Bayesian and MaxEnt approaches has been explored elsewhere[55, 56]. However, the precise connection between the MinRE metric developed here and the Bayes Factor used in the Bayesian approaches remains as an interesting topic of future work.

Technical limitations often make it difficult to perform single cell experiments due to lack of appropriate antibodies or small concentrations of expressed proteins in individual cells. In such situations immunoassays (e.g., western blot) measuring the cell population level abundances of signaling proteins are used to decipher underlying mechanisms. The MaxEnt based approach can be used in conjunction with these types of experimental assays to find the least structured distribution of protein abundances and attributes of cell responses. As our study with *E. coli* chemotaxis shows, these distributions, even after being constrained to reproduce the average values from the experimental data, may need additional constraints regarding variances and co-variances of the variables in order to reach agreement with the experimental information. However, the estimated distributions calculated using constraints on average values indicate the limits of the cell-to-cell variations that can be allowed to observe the experimental data.

As mentioned earlier, robustness and sensitivity of cell responses represent two sides of the same coin. Here we used MinRE to determine the robustness, or insensitivity of chemotactic responses in *E. coli* cells to parameter variation. However, the same method can be used to determine parameters that sensitively regulate cell functions; these parameters will vary in individual cells within a narrow range. The joint distribution of these parameters can be use to calculate covariances between the parameters. The covariance matrix analyzed using principal component analysis (PCA) can determine the most sensitive parameters or linear combinations of a subset of parameters that represent directions of sensitive perturbations[57, 58]. Therefore, the MaxEnt based analysis could be used to identify sensitive and insensitive or "sloppy" parameters[59].

In our investigations, we held the kinetic rates describing strengths of protein-protein interactions in the models fixed in individual cells. This represents a reasonable assumption in the signaling models we investigated, since kinetic rates are largely determined by the thermodynamics of protein-protein interactions. Therefore, for experiments done at a fixed temperature, very little cell-to-cell variation in the rates is expected. However, when signaling models approximate interactions between two proteins, which can be affected by molecular crowding[60, 61] and/or cell shapes, then cell-to-cell variations of such kinetic rates should be included in the calculations. Moreover, coarse-grained *in silico* models often approximate multiple steps in biochemical reactions by one-step reactions, e.g., through the Michaelis-Menten approximation, where kinetic rates of the reactions depend on protein abundances. These reaction rates could vary in *in silico* models describing signaling kinetics in individual cells due to cell-to-cell variations of protein abundances[62]. Such variations in kinetic rate constants can be easily incorporated under our approach.

In many situations, experimentally measured values of the kinetic rates of biochemical reactions and protein concentrations in cell signaling and gene regulatory systems are unavailable, especially in higher organisms. Moreover, data such as strengths of protein-protein interactions, e.g., obtained *in vitro* using truncated protein domains, may incorrectly describe those interactions *in vivo*. Therefore, an increasingly popular systems biology approach is to subject *in silico* models to perturbations of parameters (rate constants and proteins concentrations), using as a measure of robustness the insensitivity of key model outputs to such perturbations[34, 63]. For computational reasons, models with small numbers of sensitive parameters are generally preferred[63]. Our approach can also be used for analysis of *in silico* models in such settings. In the absence of any experimental data, user provided criteria for accepting a parameter set, such as production of cell population averaged concentrations of specific signaling proteins within a particular range, can be used for estimating robustness of these models.

The proposed approach is not free from some limitations of MaxEnt as a technique for inferring parameters[47, 64, 65]. First, in order to set up the MaxEnt model, one needs to start with a set of constraints that are relevant for the system. But when little is known about the biology or the functional response of a signaling network, it may be difficult to determine which are the relevant constraints from an available set of variables describing qualitative aspects of signaling kinetics. Second, it is possible for the inferred distributions of parameters to be inconsistent with experiments. This can be an informative result in its own right, since it may imply that additional constraints are required. However, increasing the number of constraints also requires solving for an increasing number of Lagrange multipliers, which can pose a significant computational challenge, in some cases requiring the use of computationally intensive Monte Carlo algorithms[66]. Since signaling networks can easily contain a large number of reactions and molecular species[67, 68], MaxEnt calculations for such networks can become computationally challenging. Third, as previously noted, the constraints used in Eqs. 4-6 refer to the expected values of variables rather than the sample averages, and in finite samples these two quantities can differ from one another, introducing error into the estimation of the Lagrange multipliers[47]. Since the Lagrange multipliers appear in the exponential functions in the estimation of $\hat{P}(\{n_q^{\text{total}}\})$, a small change in a Lagrange multiplier can potentially produce a large change in $\hat{P}(\{n_q^{\text{total}}\})$ blurring the differences between relative entropies or MinRE values for different models. However, we have evaluated the current results for robustness to address this issue, and our tests show (Supplementary Figure. S6B) that even allowing for substantial deviations between the sample averages and expected values, the rank ordering of models remains unchanged. Finally, in small samples, underlying dependencies of higher moments on lower moments could be difficult to detect. If present but not accounted for, such dependencies could lead to poor estimation of MinRE (or robustness) and a MaxEnt model with lower predictive power.

The MaxEnt based approach is general and can be used for a wide range of cell signaling systems. The ability of the approach to predict distributions of single cell attributes in a cell population makes it particularly useful in selecting the correct model when one can construct competing mechanistic models consistent with existing experimental measurements. We have observed a surge in development of new technologies for measuring signaling kinetics in single cells in the recent years. We believe the systems biology community will find the MaxEnt based approach useful for deciphering new mechanisms using single cell and cell population averaged data.

**Materials and Methods:**

**Data from *E. coli* experiments:** The average values of the chemotactic protein abundances were taken from Li et al. [40]. We considered the chemotactic response at 100µM L-aspartate stimulation. The distribution of the adaptation time was obtained from Min et al. [39] by digitizing Fig.3C in that paper using an online web plot digitizer (http://arohatgi.info/WebPlotDigitizer/). The values of $\bar{\tau}$ and $\overline{\tau^2}$ were calculated from the distribution thus obtained. The cell population level value for the precision of adaptation for wildtype RP437 strain was obtained from Alon et al [37]. In their experiments, the measured cell population averaged ratio of the steady state tumbling frequency of the wild type *E. coli* cells in the absence of any nutrient to that of when 1mM L-aspartate was added in the medium was equal to 0.98 ± 0.05. The perfect adaptation corresponds to a ratio of 1.0. We considered precision of adaptation in individual cells. For a stochastic trajectory Γ (Fig 2B), we defined the precision of adaptation $s_\Gamma$ as the absolute value of the relative difference between the population averaged abundance of pre-stimulus CheY-P at the steady state ($\bar{N}_{\text{CheY-P pre-stim}}$) and the post stimulus steady state abundance of CheY-P ($N_{\text{CheY-P post-stim}}$) in individual cells. We calculated $N_{\text{CheY-P post-stim}}$ by using the CheY-P abundance in single *E. coli* cells evaluated at t=2000s, which is about 8 times larger than the average adaptation time. The precision of adaption in a single cell is given by, $s_\Gamma = \left| \dfrac{\bar{N}_{\text{CheY-P pre-stim}} - N_{\text{CheY-P post-stim}}}{\bar{N}_{\text{CheY-P pre-stim}}} \right|$. When single *E. coli* cells adapt perfectly, $s_\Gamma \cong 0$. We calculate $\bar{s}$ from $s_\Gamma$ using Eq. 4, We varied $\bar{s}$ from 0.005 to 0.05 for Fig 3 as we have considered a concentration of 100µM L-aspartate in our simulations instead of 1mM used in Ref. [37]. For Fig 4, we have used a $\bar{s}$ value of 1- 0.98 = 0.02. In order to make the calculations computationally efficient we calculated the cell-population averaged quantity, $\bar{N}_{\text{CheY-P pre-stim}}$, by solving the ODEs which ignored the intrinsic noise of each individual *E. coli* because the contribution of the intrinsic noise fluctuation to this average value was small (Supplementary Figure 4).

The steady state abundance of CheY-P varies from cell to cell due to the variations of total protein abundances in individual *E. coli* cells. The variation of the steady state abundance of CheY-P needs to be within 30% from an optimal value for proper functioning of the flagellar motor [5]. We calculated the variation of steady state CheY-P abundance ($p$) in single *E. coli* cells using the equation below,

$$p = \left| \frac{N_{\text{CheY-P pre-stim}} - N_{\text{CheY-P optimal}}}{N_{\text{CheY-P optimal}}} \right|$$. Optimal value of CheY-P, $N_{\text{Che-Y optimal}}$, is defined as

the ODE based solution of the steady state value of the CheY-P when the total protein concentrations are set to the values quoted in Li et al [40]. $\bar{p}$ was calculated from $p_\Gamma$ using Eq. 4. For Fig 3, $\bar{p}$ has been varied from 10 to 30 % whereas for Fig. 4 we have used an ad hoc value of 20 %. We calculated $N_{\text{Che-Y pre-stim}}$ using the ODE solutions ignoring intrinsic noise fluctuations for the reasons mentioned above.

The single cell CheY distribution for the wildtype RP437 strain was taken from Kollman et al [5]. The plot was digitized using the web plot digitizer and the x-axis was rescaled to obtain the average number of the CheY protein quoted in [40]. The moments for abundances of CheY were calculated using this distribution. The moments for CheZ abundances were computed by digitizing the co-expression plot in [5].

**Computational Method:** We use a rule based modeling software package BioNetGen[68] for the solution of the ODEs as well as to perform continuous time Monte Carlo. Using the BioNetGen software for simulating the models allows us to share our codes easily, in addition to varying the parameters efficiently in the simulations. The codes are available at http://planetx.nationwidechildrens.org/~jayajit/. The biochemical networks for the BL and the MBL models are curated from Ref. [5]. The FT model is constructed by adding an extra module to the MBL model where CheB-P is allowed to de-phosphorylate the inactive receptors. All the simulations were initialized at $t^{\text{initial}} = -800000$ s with protein abundances, $\text{Tar}_{m=0}(t^{\text{initial}}) = \text{Tar}^T$, $\text{CheA}(t^{\text{initial}}) = \text{CheA}^T$, $\text{CheR}(t^{\text{initial}}) = \text{CheR}(t) = \text{CheR}^T$, $\text{CheB}(t^{\text{initial}}) = \text{CheB}^T$, $\text{CheY}(t^{\text{initial}}) = \text{CheY}^T$ and $\text{CheZ}(t^{\text{initial}}) = \text{CheZ}(t) = \text{CheZ}^T$ while abundances of all other species (methylated receptors ($m \neq 0$) and phosphorylated form of all the other proteins) were set to zero. The superscript T refers to total abundances of the respective protein in a single *E. coli* cell. The non-zero protein abundances at $t^{\text{initial}}$ are drawn from a uniform distribution $U(0, U_H)$, where $U_H$ is chosen to be roughly 10 times larger than the experimentally measured mean abundance of the corresponding chemotactic protein [40] (see supplementary Table 2 for the mean values used in our MaxEnt calculations). Once a set of initial protein abundances is chosen, we solve the ODEs describing the chemotactic kinetics to obtain steady state values of the protein abundances. These steady state values for the protein abundances are used as initial conditions to simulate the chemotactic response of individual cells when 100μM of L-aspartate is added to the medium. The time when the ligand is added is considered to be $t=0$. The chemotactic response of the *E. coli*, given by the kinetic of the protein abundances, is calculated by solving the Master Equation in Eq. 3 exactly by the Gillespie method. The ligand receptor interactions are approximated by the rates at which the receptors can become active. The adaptation time ($\tau_\Gamma$) and the precision of adaptation ($s_\Gamma$) are calculated using for each stochastic trajectory $\Gamma$ representing the chemotactic response in an individual cell composed of abundances of signaling proteins recorded at regular time intervals for a time period of $t_0=0$ to $t_n=2000s$; the time scale, $t_n=2000$ is much larger than the typical adaptation time for E.coli for a 100μM L-aspartate stimulation. When the CheY-P abundance in an individual cell does not recover to the half of the pre-stimulus CheY-P level within 2000 s, we assign a very number (6 x$10^6$ s) to $\tau$ to mark the cell that did not adapt in a realistic time scale. In size of the sample (up to 70,000 single E. coli cells), each individual E. coli cell produced a unique stochastic trajectory. Therefore, in our simulations each stochastic trajectory could be identified with a single cell.

**Calculation of MinRE:** We seek for a solution of the form given in Eq. (9), when the average values and the nth moments of the chemotactic responses { $f_{a\Gamma}$ } and the protein abundances { $n_q^{\text{total}}$ } are constrained to the value measured in the experiments. The Lagrange multipliers { $\lambda_a$ }, { $\kappa_a$ }, { $\mu_q$ }, and, { $\eta_q$ } are obtained by solving the set of nonlinear simultaneous equations when Eq. (9) is substituted in the equations (Eqns. 4-6) describing the constraints. We carry out the summation over the stochastic trajectories { $\Gamma_C$ } then evaluate average values in the constraint equation in the following way. In the sample size we considered, each *E. coli* cell produces a unique chemotactic response composed of a stochastic trajectory describing time evolution of abundances of signaling proteins, therefore, we identified each trajectory by the single cell that generated it (Fig. 2). The single cells were assigned with identification numbers, such as cell #1, cell#2 and so on. The summation over the trajectories then is essentially the summation over the single cells used in the simulation. The uniqueness of the stochastic trajectories in our simulations also implies that for a particular stochastic trajectory, $\Gamma'_C = \{\{n'_j\}, t_n ; \{n'_j\}, t_{n-1} ; ....; \{n'_j\}, t_0\}$ for a fixed set of total protein concentrations { $n'^{\text{total}}_q$ }, $P_C = P(\{n'_j\}, t_n ; \{n'_j\}, t_{n-1} ; ....; \{n'_j\}, t_0 | \{n'^{\text{total}}_q\})$ is either equal to 1 (when $\Gamma'_C$ or the corresponding single cell is present in the samples we considered) or 0 (when $\Gamma'_C$ is absent in the samples). Thus, $\sum_{\{\Gamma_C\}} P_C \ln P_C = 0$ and $\sum_{\{\Gamma_C\}} f_{a\Gamma} P_C = f_{a\Gamma_C}$ for the trajectories we analyzed in the simulations. We used up to 70, 000 single cells, the convergence of the results with the number of cells used is shown in the supplementary material (Supplementary Figure 5). The Lagrange multipliers are then calculated using standard techniques used for solving non-linear algebraic equations (web supplement). Then we calculate the minimum relative entropy MinRE given by Eq. (8). It is in principle possible that when a very large number of cells are present, two different single cells could produce the same stochastic trajectory and for such cases, $\sum_{\{\Gamma_C\}} P_C \ln P_C$ will not vanish and $\sum_{\{\Gamma_C\}} f_{a\Gamma} P_C$ will contain averages over multiple trajectories. However, occurrences of such events (e.g., the presence of pairs of identical stochastic trajectories) appear to be extremely rare for the rate constants and the ranges of the protein abundances we considered. We further tested this approximation by considering deterministic chemotactic signaling kinetics where the kinetics of signaling protein abundances only depend on the total protein abundances (as the kinetic rates are fixed for each cell), therefore, $P_C=1$ when the deterministic kinetic trajectory of abundances of signaling proteins is present and $P_C=0$, otherwise. When we used the same a priori uniform distribution for protein abundances as our stochastic simulations, the rank ordering of the FT, BL, and the MBL models based on the MinRE values did not change compared to the stochastic simulations (Supplementary Figure. S7). The small differences in the values of the MinRE between the stochastic and the deterministic simulations show the dominance of extrinsic noise fluctuations over intrinsic noise fluctuations in determining robustness of the models (Supplementary Figure. S7). These results also demonstrate that associating a unique stochastic trajectory to a single cell is a good approximation for the calculation of MinRE in the stochastic simulations.

**Prediction of single cell properties:** Our MaxEnt approach allows us to predict distributions of single cell properties given by Eq. (9). The Lagrange multipliers in Eq. 9 are calculated using the procedure described above. Once the Lagrange multipliers are known, Eq. 9 essentially predicts the probability for an individual cell (indexed by the stochastic trajectory Γ) displaying chemotactic responses { $f_{a\Gamma}$ } and the protein abundances { $n_q^{\text{total}}$ }. We use this probability to calculate occurrence probabilities for the cells in the cell population that were initially assigned with protein abundances chosen from uniform distributions. The occurrence probability evaluated for an individual cell gives the maximally broad probability with which that particular cell should be present in the cell population so that the cell population is able to produce the population level and single cell measurements that were constrained in the MaxEnt calculation. For example, when $\bar{\tau}$ is constrained to its experimentally measured value $\bar{\tau}^{\text{expt}}$, then the MaxEnt calculation produces a solution
$\widehat{P}_\Gamma = Z^{-1} \exp\left[-\lambda_\tau \tau_\Gamma\right]$, where, $\tau_\Gamma$, is the adaptation time for an individual cell executing a stochastic trajectory, Γ, and, $\lambda_\tau$ is the Lagrange multiplier. Once, $\lambda_\tau$ is evaluated using $\bar{\tau}^{\text{expt}}$, $\widehat{P_\Gamma}$ gives the probability that an individual cell, $\Gamma$, is selected in the cell population. The relationship between $\widehat{P_\Gamma}$ and $p(\tau)$ is given by Eqn (1). Following the same scheme, the other single cell properties, such as CheY abundance, or, CheZ abundance are evaluated. The moments of the distributions are then calculated using those distributions.

**ACKNOWLEDGEMENTS**
This work was supported by funding from the Research Institute at Nationwide Children's Hospital and NIH grant AI090115 to J.D., and NIH grant MH086117 to V.J.V. J.D. thanks C. Jayaprakash for discussions. We thank both the anonymous reviewers for making constructive suggestions.

**Authors Contribution**

JD, SM and VJV planned the research and analyzed the data; SM and SS performed simulations and calculations. JD, SM and VJV wrote the paper.

**Conflict of Interest**

We declare no conflict of interest.

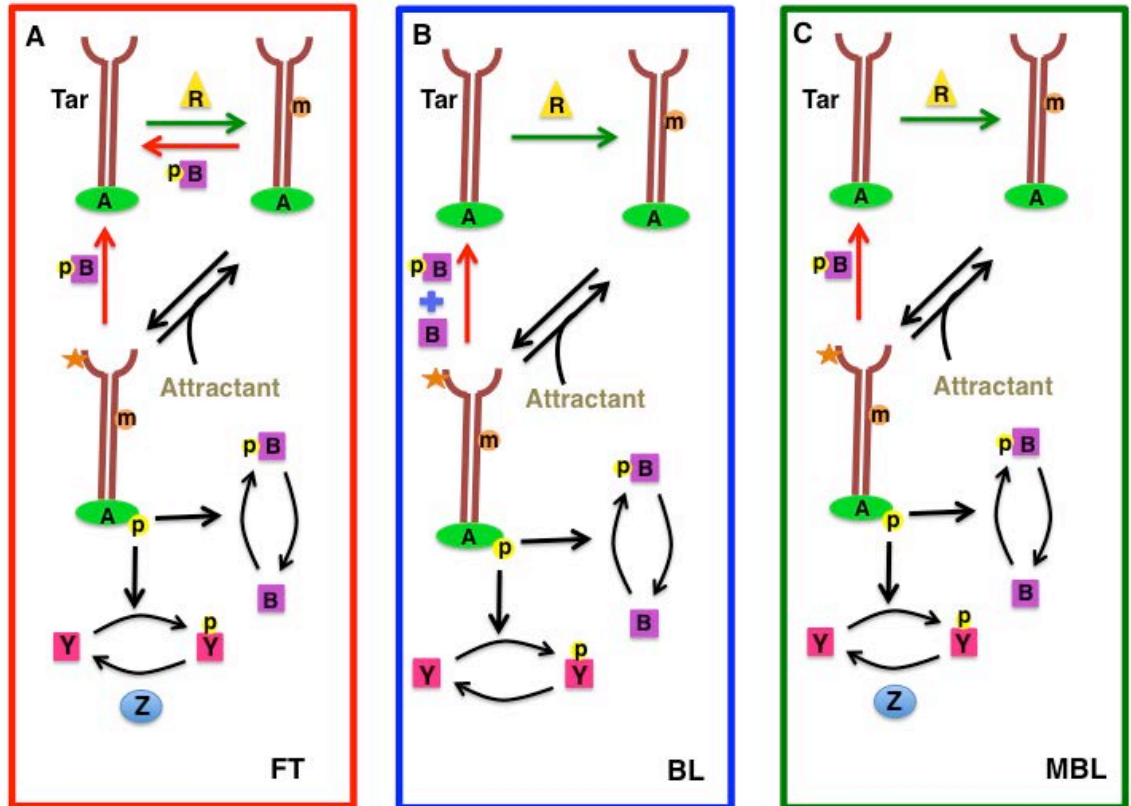

**Fig. 1. Three models for *E. coli* chemotaxis. (A)** The Fine tuned model (FT), originally proposed by Knox et al [42] and Hauri and Ross [41], shows robust adaptation only for a small region in the parameter space. In this model the aspartate receptors (Tar) can shuttle between an active (denoted by the orange star) and an inactive conformation. The probability of the receptors to be in an active conformation depends on their state of methylation. Pair of enzymes namely CheR (R) and CheB-P (B) add and remove methyl groups from the receptors. CheB-P can demethylate receptors regardless of *their state of activity*. The Tar receptors form complexes with a kinase CheA (A). CheA can phosphorylate itself with a rate proportional to the abundance of the active Tar. Phosphorylated CheA can transfer its phosphoryl group ($PO_4$) either to the kinase CheB, rendering it capable of demethylation, or to another response regulatory protein called CheY (Y). Upon receiving the phosphoryl group from CheA, CheY renders itself active. Active form of CheY (CheY-P) diffuses across the cell and binds to the flagella motors causing them to tumble. The phosphatase CheZ (Z) de-activates the active form CheY.
**(B)** Barkai Leibler model (BL) was put forward to explain robust adaptation in the chemotactic network of bacterial *E. coli*. Unlike FT model, CheB, both the unphosphorylated and the phosphorylated form, demethylates only the *active* Tar receptors. This model also lacks the phosphatase CheZ and CheY-P undergoes auto dephosphorylation. **(C)** MBL model proposed by Kollmann et al., except for two differences, is similar to the BL model. The differences are: i) Only the *phosphorylated* form of CheB, as opposed to CheB and CheB-P in the BL model, can de-methylate the active receptors. ii) the dephosphorylation of the active CheY is done by the phosphatase called CheZ (Z). The presence of this phosphatase makes the steady state of CheY-P abundances relatively robust to the concerted over expression of the chemotactic proteins.

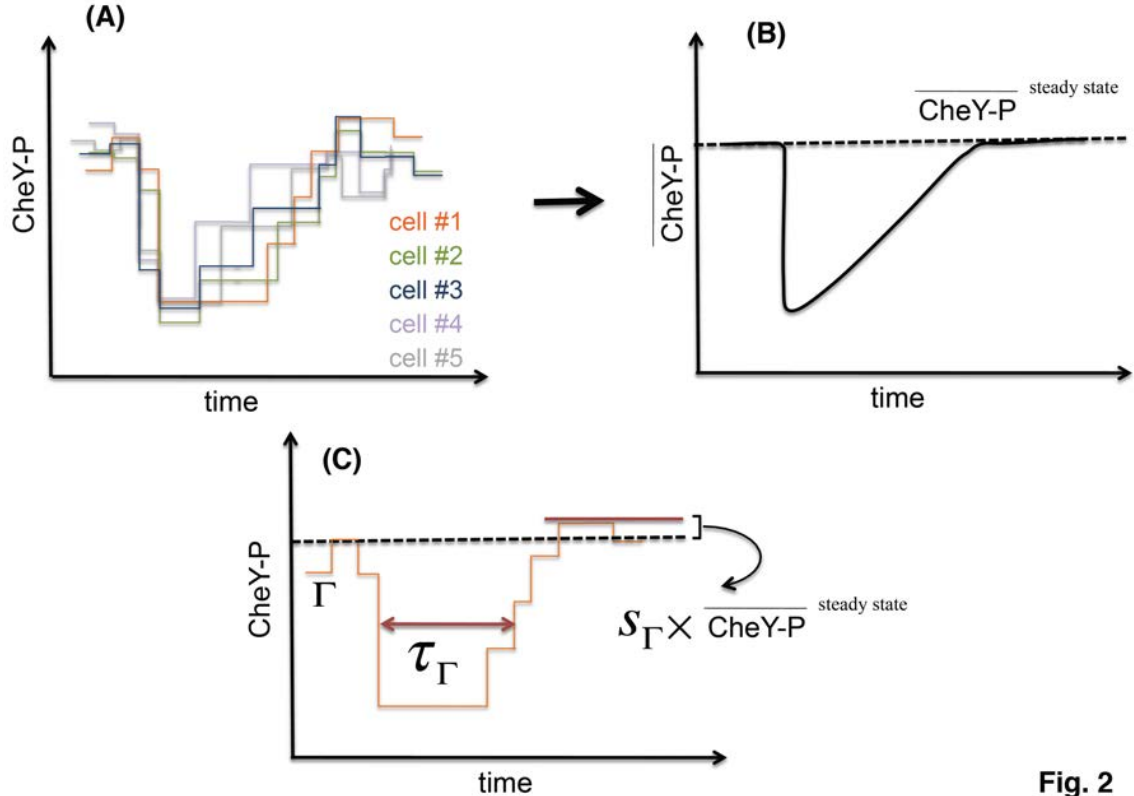

**Fig.2. *E. coli* chemotactic response in individual cells and in a cell population. (A)** Schematic diagram shows kinetics of CheY-P abundance in individual E. coli cells (indicated with different colors) when attractants are added in the medium at $t = t_0$. The temporal profile of CheY-P abundance varies from cell-to-cell due to intrinsic and extrinsic noise fluctuations in the signaling kinetics. **(B)** Shows schematically kinetics of cell population averaged concentration of CheY-P. The dashed line displays the pre-stimulus steady state concentration ($\overline{\text{CheY-P}}^{\text{steady state}}$) of CheY-P in a cell population. Experiments such as immunoblot assays measure such cell population averaged kinetics of signaling kinetics. **(C)** Each individual cell produces a temporal profile or a stochastic trajectory, $\Gamma$, as described in the main text. The steady state population averaged concentration of CheY-P at the pre-stimulus level as in (B) is shown by the dashed line. We use $\Gamma$ to identify an individual *E. coli* cell, and, properties of chemotactic responses for that individual cell such as the adaptation time ($\tau_\Gamma$) or the precision of adaptation ($s_\Gamma$) can be calculated from the stochastic trajectory, $\Gamma$.

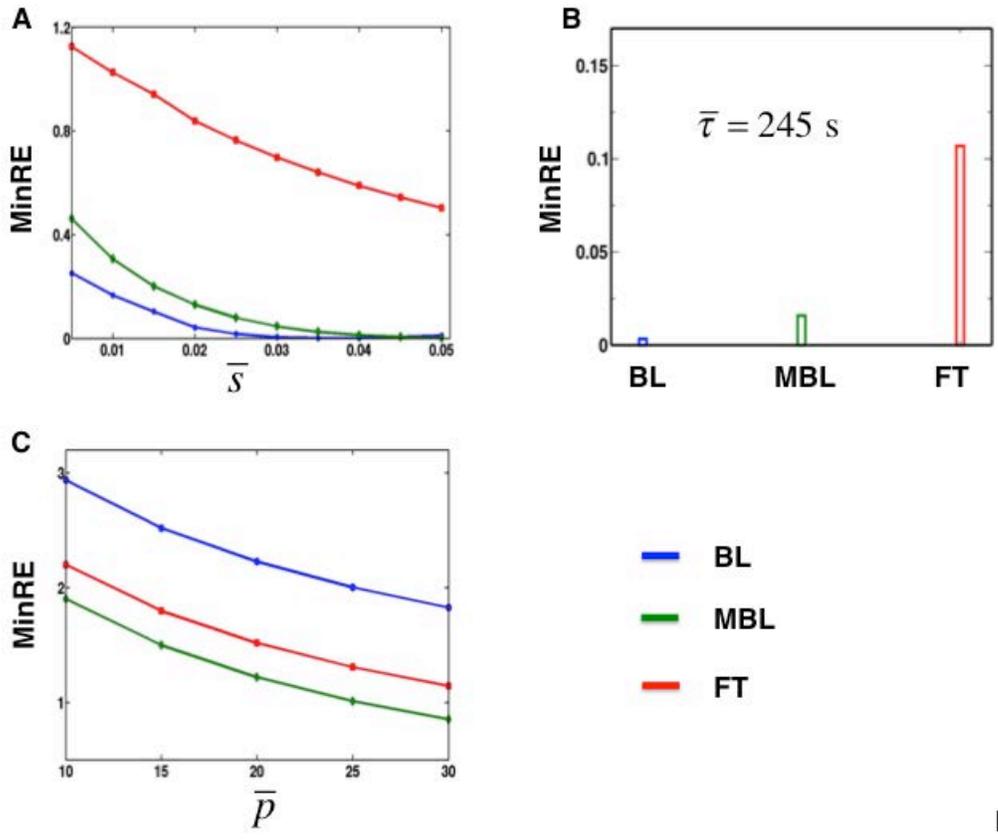

Fig. 3

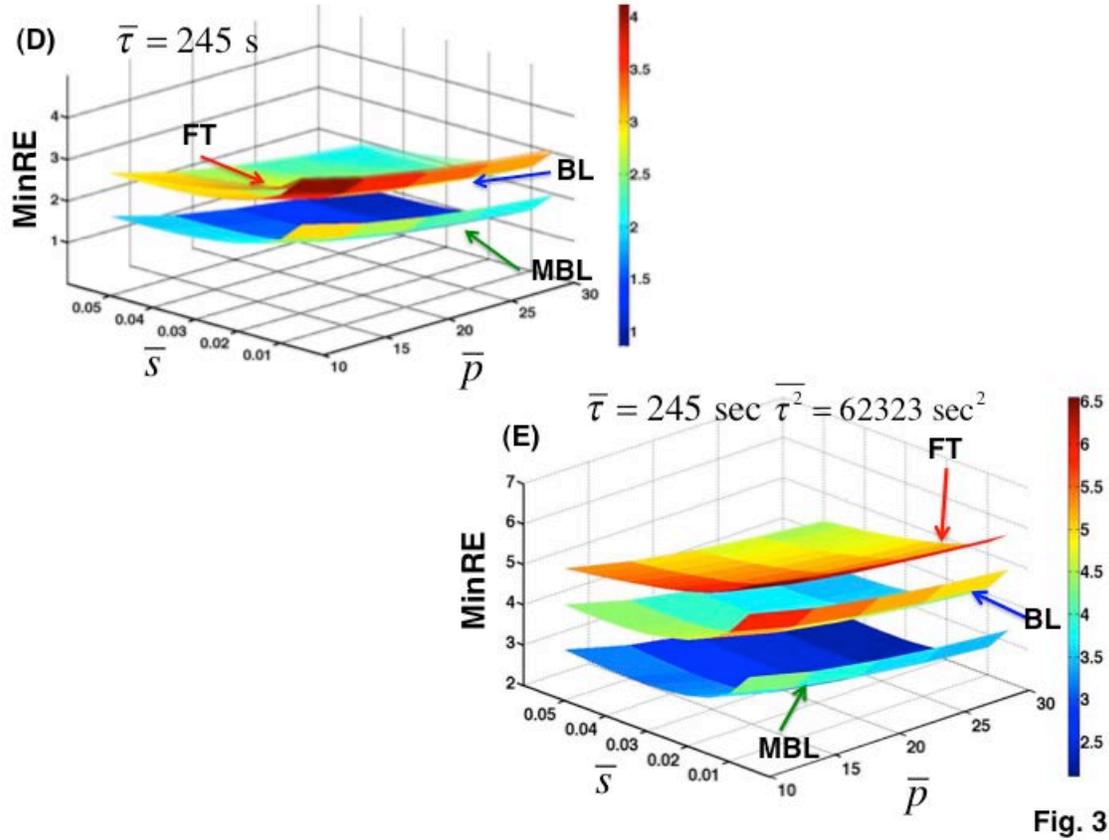

**Fig. 3. MinRE quantifies robustness of different models for *E. coli* chemotaxis. (A)** Variation of MinRE with the average precision of adaptation $\bar{s}$ for three different models (BL in blue, MBL in green and FT in red). **(B)** MinRE for the three different models when the average time is constrained to $\bar{\tau}$ =245s. The color scheme is same as (A). **(C)** Variation of MinRE with $\bar{p}$ for three different models. The same color scheme as (A) has been used. **(D)** Shows variation of MinRE with $\bar{s}$, and, $\bar{p}$ when $\bar{\tau}$ is held fixed to the experimentally measured value $\bar{\tau}^{\text{expt}}$ (=245s). **(E)** Variation of MinRE with $\bar{s}$, and, $\bar{p}$ when $\bar{\tau}$ and $\overline{\tau^2}$ are set equal to their experimentally measured values, $\bar{\tau}^{\text{expt}}$ (=245s) and $\overline{\tau^2}^{\text{expt}}$ (= 62323.5s$^2$), respectively.

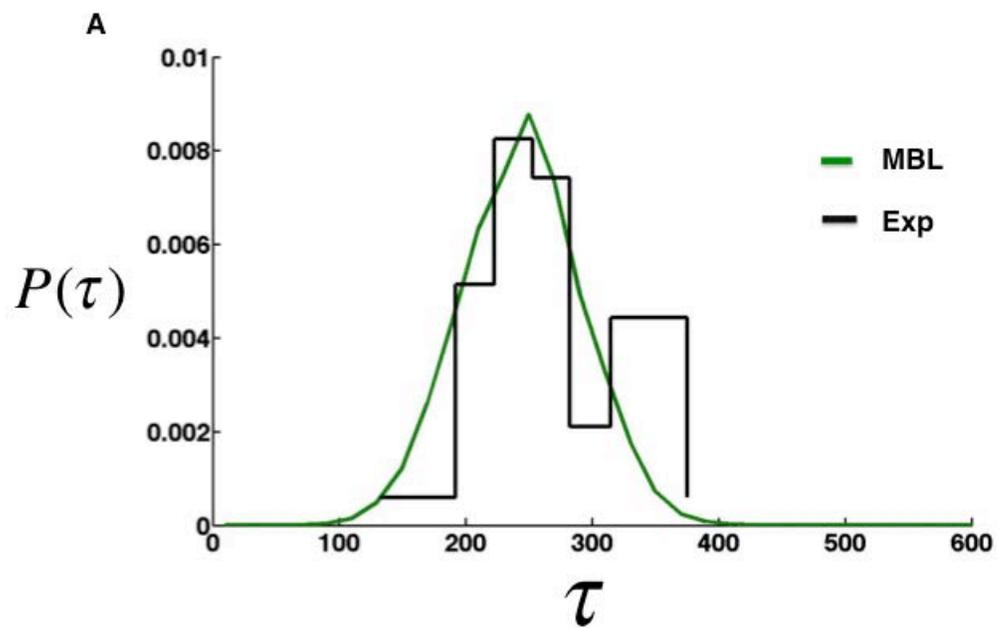

Fig. 4

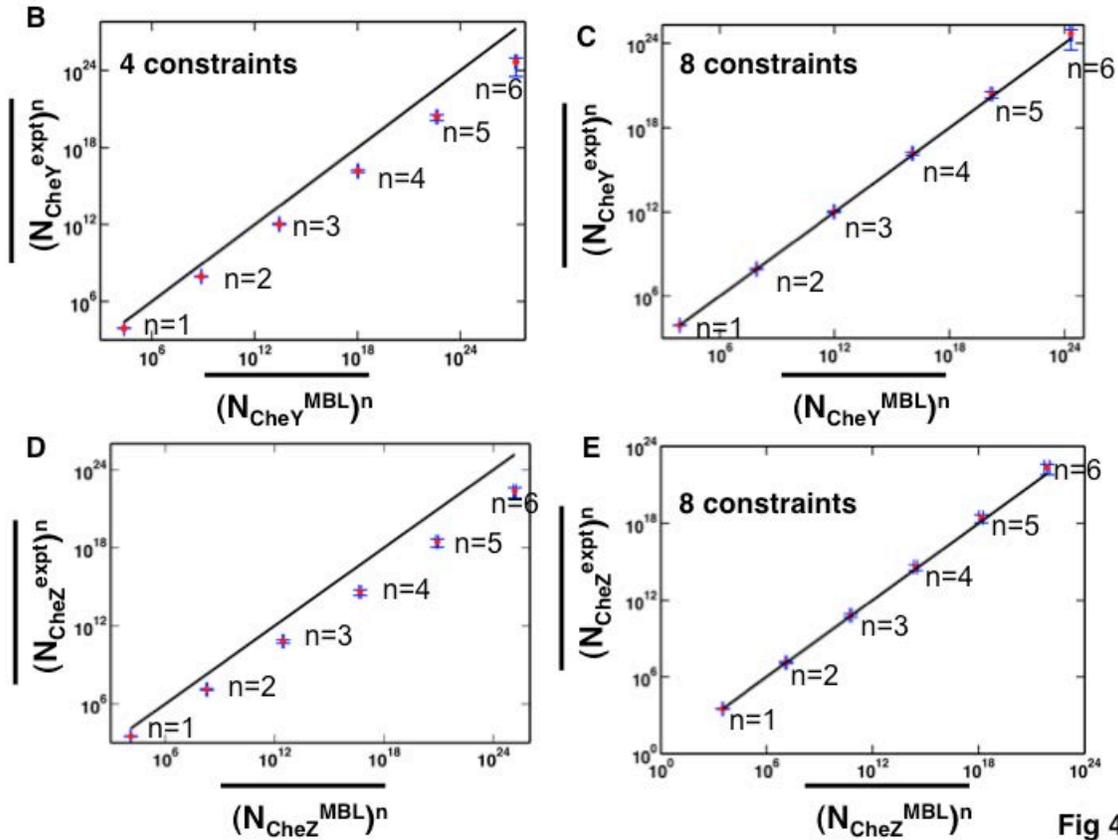

**Fig. 4. Comparison of single cell attributes as predicted from the MaxEnt approach with experiments.** (A) Comparison of the distribution of the adaptation time $\tau$ calculated from the MaxEnt distribution with the four constraints ($\bar{\tau}$ = 245 s, $\overline{\tau^2}$ = 62323 s$^2$, $\bar{s}$ = 0.02 and $\bar{p}$ = 20% ) for the MBL model of *E. coli* chemotaxis and with experiments (black stairs). (B) Comparison of the first six moments calculated from the distribution of CheY abundance obtained from the MaxEnt approach (x axis, Log scale) for the MBL model with the same moments obtained from experiments (y axis, Log scale) as reported in Ref.[5]. The MaxEnt calculation was performed using the same constraints as in (A). The black line shows the y=x graph. (C) Similar comparison between the MaxEnt predictions and experiments as in (B) for the moments calculated for the distribution of CheY. The MaxEnt approach used three constraints ($\bar{\tau}$ =245s, $\bar{s}$ = 0.02 and $\bar{p}$ = 20% ) along with the average values, variances and co-variances of abundances of CheY and CheZ obtained from experiments. The averages for the CheY abundance were obtained from Li et al.[40] and the variances of CheY and CheZ abundances and co-variance between CheY and CheZ abundances were taken from Kollmann et al. [5]. (D) Shows comparison between predictions (x axis, Log scale) generated from the MaxEnt approach and experiments (y axis, Log scale) for the first six moments calculated from the distribution of CheZ abundances in single *E. coli* cells. The distribution of CheZ abundances in the MaxEnt approach is calculated for the same constraints as in (A). The black line shows the y=x graph as a guide. The same sources as in (B) are used to obtain the experimental data. **(E)** Same comparison as in (D). The distribution of the CheZ abundance in the MaxEnt approach is calculated using the same constraints as in (C). The same sources as in (C) are used to obtain the experimental data.

Supplementary Material for "Data-driven quantification of robustness and sensitivity of cell signaling networks"

This File contains

1. The three models of E. coli chemotaxis.
2. Supplementary Table 1: Extra module in FT model
3. Supplementary Table 2: Mean values of the chemotactic proteins
4. Supplementary Table 3: Constraints used in Figure 4
5. Supplementary Table 4: Test of dependency of the variances to the average values for the constraints
6. Discussion on the adaptation time of E. coli and the role of CheB
7. Supplementary Figure 1. A simple model of exact adaptation
8. Supplementary Figure 2. The two constraints MinRE for the BL, MBL and the FT models.
9. Supplementary Figure 3. Distribution of adaptation time with two different sets of output   constraints.
10. Supplementary Figure 4. The role of intrinsic noise at pre-stimulus steady state.
11. Supplementary Figure 5. Plot for the convergence of MinRE
12. Supplementary Figure 6. Plot for extra constraints and sensitivity of MinRE towards the constraints used.
13. Supplementary Figure 7. The role of intrinsic noise in calculation of MinRE.
14. Derivation of Eq. 9 in the main text.

## The FT, BL and the MBL model of *E.coli* chemotaxis

The Barkai-Leibler (BL) and the modified Barkai-Leibler (MBL) models have been adopted from Kollmann et al (1). The rate constants and the relevant reactions can be found in their Supplementary Material. In order to facilitate comparison across different models, our FT model was constructed just by adding an extra module to the MBL model where the CheB-P can de-methylate the inactive forms of the Tar receptor complex ($Tar_m$ as opposed to $Tar_m^A$) as well. The Michaelis-Menten rate constants are given below.

Supplementary **Table S1: Extra module in FT model**

| CheB-P+$Tar_m$ $\rightleftarrows$ CheB-P-$Tar_m$ $\rightarrow$ CheB-P+$Tar_{m-1}$ | Michaelis-Menten Constant $K_b$=2.5 μM | Catalytic rate $k_b$ = 6.3 $s^{-1}$ |
|---|---|---|

Our FT model, though very similar in essence to the original Knox et al. (2) and Hauri and Ross (3) model, harbors a few differences worth mentioning. Hauri and Ross approximated the Tar, CheW and CheA into one complex, which they referred to as T. They did not consider an explicit auto phosphorylation of CheA. We have an explicit auto phosphorylation reaction of CheA with a rate proportional to the total number of active Tar complex. Unlike Hauri and Ross model, CheA instead of the complex T transfers the phosphate group to CheY and CheB. We also have explicitly considered the phosphatase CheZ. Hauri and Ross assumed a first order de-activation of CheY-P.

Supplementary **Table S2: Mean values of the chemotactic proteins**

| Proteins | Mean Experiment (4) (in units of # of molecules in a single cell) | Mean used in the MaxEnt approach (in units of # of molecules in a single cell) |
|---|---|---|
| Tar + Tsr | 15000 ± 1700 | 15000 |
| CheA | 4452 ± 920 | 4452 |
| CheY | 8148 ± 310 | 8148 |
| CheR | 140 ± 10 | 134 |
| CheB | 240 ± 10 | 235 |
| CheZ | 3200 ± 90 | 3192 |

Supplementary **Table 3: Constraints used in Figure 4**

| Constraints | Experiments | Value used |
|---|---|---|
| $\bar{\tau}$ | 245 ± 17.5274 s (5) | 245 s |
| $\bar{s}$ | 0.02 ± 0.05 (6) | 0.02 |
| $\bar{p}$ | < 30% (1) | 20% |
| $\overline{\tau^2}$ | 62323.5 ± 9025.87 s² (5) | 62323 s² |
| $\overline{CheY}$ | 8148 ± 310 (4) molecules/cell | 8148 molecules/cell |
| $\overline{CheZ}$ | 3200 ± 90 (4) molecules/cell | 3192 molecules/cell |
| $\overline{CheY^2}$ | 85581400 ± 6097680 (1) molecules²/cell | 82987380 molecules²/cell |
| $\overline{CheZ^2}$ | 12914000 ± 2035600 (1) molecules²/cell | 12736080 molecules²/cell |
| $\overline{CheY.CheZ}$ | 31942100 ± 5302080 (1) molecules²/cell | 32510520 molecules²/cell |

Supplementary **Table 4: Test of dependency of the variances to the average values for the constraints using $\chi^2$**.

| Variable | Exponential | Poisson | Normal |
|---|---|---|---|
| CheY abundance | 4427.79 | 38.9494 | 0.913494 |
| CheZ abundance | 2490.89 | 11.1413 | 1.66981 |
| Adap time (τ) | 795083 | 1.51498 | 0.0024179 |

The exponential and the Poisson distributions were generated from the mean values of variables. The exponential distribution for the # of CheY molecules, for example, is given by $e^{-x/\overline{CheY}} / \sum_{CheY} e^{-x/\overline{CheY}}$ where $\overline{CheY}$ is given in Supplementary Table 2. Similarly, for the Poisson distribution of the form $(e^{-\lambda}\lambda^x)/x!$, we set λ to be equal to the average values of the variables under consideration. If indeed the experimental distribution resembles either the exponential distribution or the Poisson distribution, the mean alone would determine all the higher order moments.

We calculated the mean and the variance for the normal distribution using the mean and the standard deviations calculated from the experimental data. Then we generated higher moments (up to 6) from the generated exponential, Poisson, or the normal distribution and calculated the $\chi^2$ using the formula below

$$\chi^2 = \sum_{n=1}^{6} \frac{\left(\overline{(Y^{expt})^n} - \overline{(Y^{dist})^n}\right)^2}{\left(\sigma_{expt}^{(n)}\right)^2}$$, where $\overline{(Y^{expt})^n}$ is the n$^{th}$ moment of the experimental distribution and $\overline{(Y^{dist})^n}$ is the n$^{th}$ moment of the assumed (i.e. exponential, Poisson or normal) distributions. $\sigma_{expt}^{(n)}$ is the error in the measurement of the n$^{th}$ experimental moment. The values of the $\chi^2$ for different distributions are shown in the above table.

### Time scales of adaptation

A simple model, taken from Alon et al. (7), (see Fig S1) has been analyzed here. Instead of several methylation sites, the receptors are assumed to have only one site for methylation. The methylation is done by the kinase CheR, which works at saturation. A receptor with no methylation is permanently inactive while a methylated receptor can become active at a rate that depends on the ligand concentration. The kinase CheB works only on the active form of the methylated receptor. Instead of an enzymatic de-activation, we have assumed a first order de-activation of the active receptors for analytic simplicity. The ODEs are given by

$$\frac{dX_0}{dt} = V_B B X_1^* - V_R R$$

$$\frac{dX_1}{dt} = V_R R - k_1 X_1 + k_{-1} X_1^*$$

$$\frac{dX_1^*}{dt} = k_1 X_1 - k_{-1} X_1^* - V_B B X_1^* \qquad (1)$$

where $X_0$, $X_1$ and $X_1^*$ are the de-methylated, methylated and active methylated receptors respectively. $B$ and $R$ stand for the concentrations of CheB and CheR. $V_B$ and $V_R$ are the de-methylation and the methylation rates. The methylated receptor $X_1$ can become active at a rate $\alpha = k_1/(k_1+k_{-1})$. This rate $\alpha$ depends on the ligand concentration L. Addition of nutrients make $\alpha$ plummet at a time scale much faster to the time scales of methylation and de-methylation. We can see from Eqn (1) however, that regardless of what the value of $\alpha$ is, $X_1^*$ adapts to the same steady state value of $(X_1^*)_s = V_R R/V_B B$. The node $X_0$ works to integrate the error in adaptation and feed the integrated error back to the input as we can see from Eqn (2), guaranteeing perfect adaptation.

$$\frac{dX_0}{dt} = V_B B \left( X_1^* - \frac{V_R R}{V_B B} \right) = V_B B \left( X_1^* - (X_1^*)_s \right) \qquad (2)$$

We will now proceed to estimate the speed of adaptation of the module described above. Owing to the conservation law $X_0 + X_1 + X_1^* = \beta$, it will suffice just to focus on the time evolution of $X_1$ and $X_1^*$. Therefore,

$$d_t \begin{pmatrix} X_1 \\ X_1^* \end{pmatrix} = \begin{pmatrix} V_R R \\ 0 \end{pmatrix} + \begin{pmatrix} -k_1 & k_{-1} \\ k_1 & -(k_{-1}+V_B B) \end{pmatrix} \begin{pmatrix} X_1 \\ X_1^* \end{pmatrix} \qquad (3)$$

Defining a linear transformation as $\tilde{X}_1 = X_1 + \chi$ and $\tilde{X}_1^* = X_1^* + \varphi$, where $\chi = -V_R R(k_{-1}+V_B B)/k_1 V_B B$ and $\varphi = -V_R R/V_B B$, Eqn (3) can be recast as

$$d_t \begin{pmatrix} \tilde{X}_1 \\ \tilde{X}_1^* \end{pmatrix} = \begin{pmatrix} -k_1 & k_{-1} \\ k_1 & -(k_{-1} + V_B B) \end{pmatrix} \begin{pmatrix} \tilde{X}_1 \\ \tilde{X}_1^* \end{pmatrix} = \mathbb{M} \begin{pmatrix} \tilde{X}_1 \\ \tilde{X}_1^* \end{pmatrix}$$

The eigen values of the matrix $\mathbb{M}$ are

$$\begin{aligned}\lambda_{\pm} &= \frac{1}{2}\left\{-(V_B B + k_1 + k_{-1}) \pm \sqrt{(V_B B + k_1 + k_{-1})^2 - 4k_1 V_B B}\right\} \\ &= \frac{1}{2}\left\{-(V_B B + k_1 + k_{-1}) \pm (V_B B + k_1 + k_{-1})\sqrt{1 - \frac{4k_1 V_B B}{(V_B B + k_1 + k_{-1})^2}}\right\}\end{aligned} \quad (4)$$

Therefore to O($\varepsilon$) where $\varepsilon = 4k_1 V_B B / (V_B B + k_1 + k_{-1})^2$, we have

$\lambda_+ = -\dfrac{k_1 V_B B}{V_B B + k_1 + k_{-1}}$ and $\lambda_- = -(V_B B + k_1 + k_{-1}) + \dfrac{k_1 V_B B}{V_B B + k_1 + k_{-1}}$. Now the time scales of ligand-induced modifications are much faster compared to the time scales methylation and de-methylation i.e. $V_B B < k_1, k_{-1}$. Therefore,

$$\lambda_+ = -\alpha V_B B$$
$$\lambda_- = -V_B B (1 - \alpha) - (k_1 + k_{-1}) \quad (5)$$

So we see that

$\begin{pmatrix} \tilde{X}_1 \\ \tilde{X}_1^* \end{pmatrix} = \mathbb{C} \begin{pmatrix} e^{\lambda_+ t} \\ e^{\lambda_- t} \end{pmatrix}$, where the matrix $\mathbb{C}$ is determined by the initial conditions. The two relevant time scales in the problem are $\lambda_+$ and $\lambda_-$ respectively, both of which depend on the absolutely value of CheB (B) for a given concentration of ligand L. So if the number of CheB gets very small the module described above will take longer to adapt.

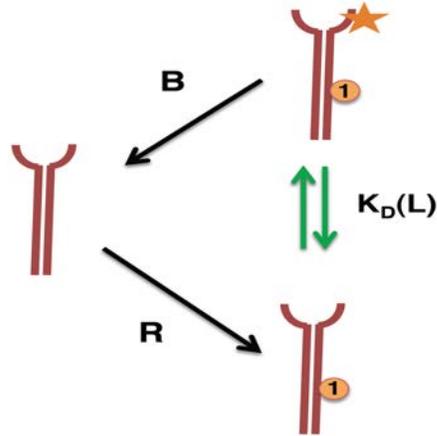

**Supplementary Figure 1: A simple model of exact adaptation:** The receptor has only one methylation site. The receptors devoid of any methylation are always inactive. The singly methylated receptors shuttle between active and inactive state with a rate $\alpha = k_1/(k_1+k_{-1})$. The rate $\alpha$ depends on the ligand concentration (L). When attractants like L-aspartate are added, $\alpha$ value decreases. The kinase CheB (B) removes methyl group only from the active receptors (denoted by orange stars) while the kinase CheR (R) adds methyl group to the de-methylated receptors.

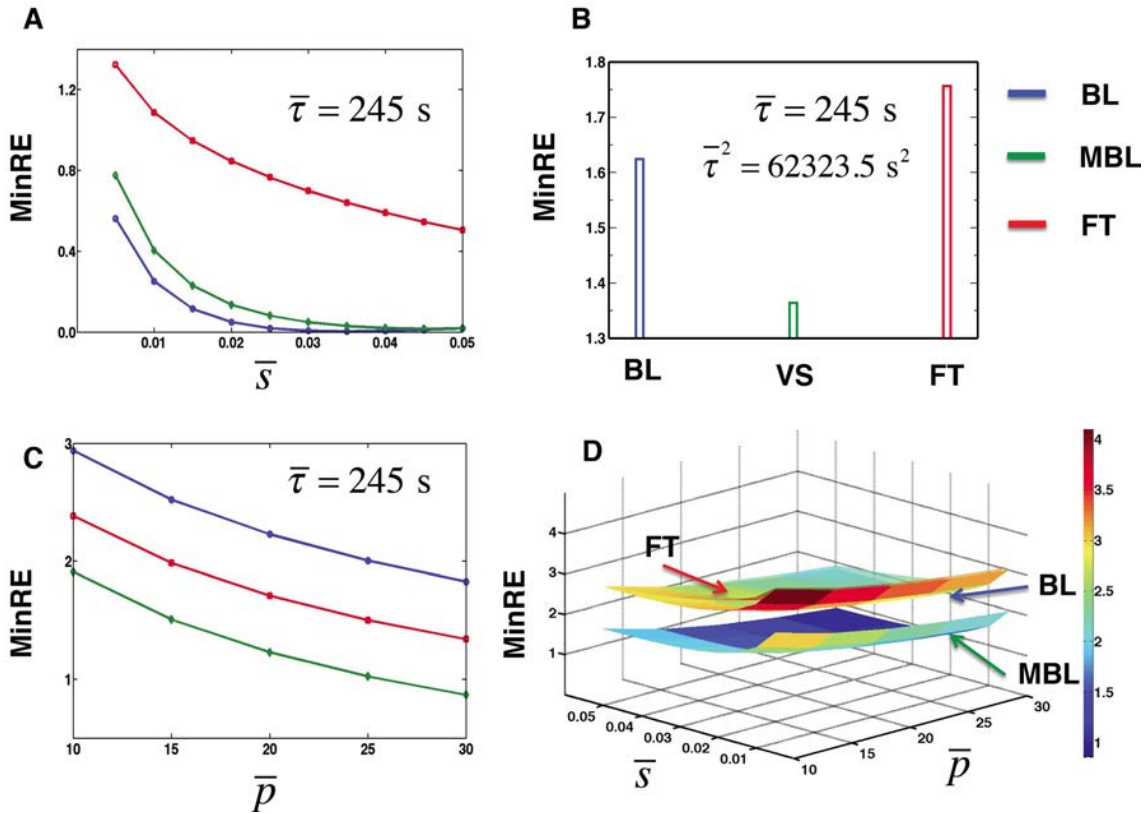

**Supplementary Figure 2: The two constraints MinRE for the three different models:** (A) The MinRE for the BL (blue), MBL (green) and the FT (red) models when the average adaptation time $\bar{\tau}$ is constrained to 245 s and the average precision of adaptation $\bar{s}$ is varied from 0.005 to 0.05. (B) The MinRE for the BL (blue), MBL (green) and the FT (red) models when the average adaptation time $\bar{\tau}$ and its square average $\overline{\tau^2}$ are constrained to the values of 245 s and 62323.5 s². (C) The MinRE for the BL (blue), MBL (green) and the FT (red) models when the average adaptation time $\bar{\tau}$ is constrained to 245 s and the average percentage of variation $\bar{p}$ is varied from 10 to 30 %. (D) A 2D surface plot of MinRE for three different models (shown with the colored arrows) when the average precision of adaptation $\bar{s}$ and the average percentage of variation $\bar{p}$ are varied from 0.005 to 0.05 and from 10 to 30 % respectively.

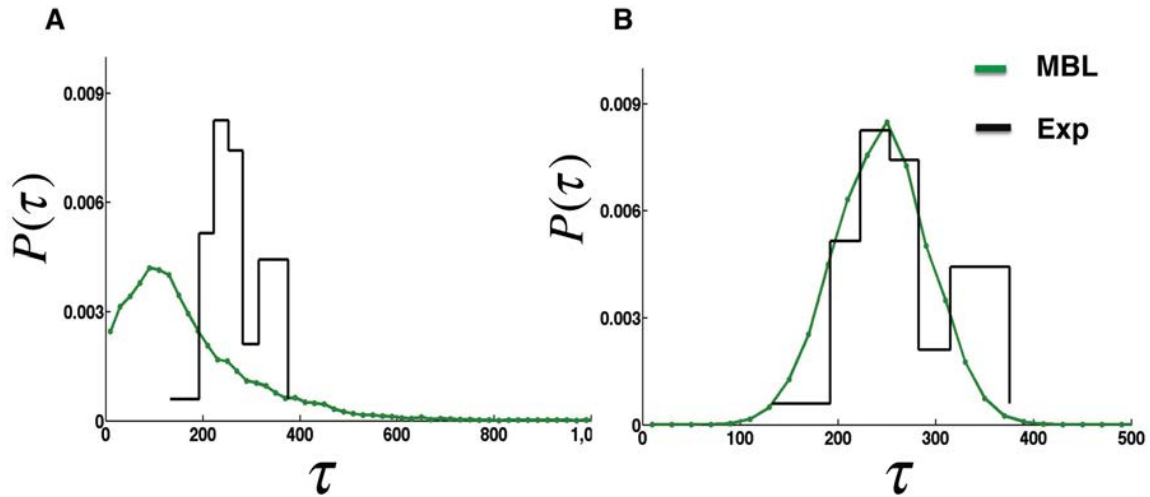

**Supplementary Figure 3: Distribution of adaptation time with two different sets of output constraints:** (A) The adaptation time distribution for MBL (green) model extracted from the MaxEnt distribution when the three output constraints namely $\bar{\tau}$, $\bar{s}$ and $\bar{p}$ are constrained to 245 s, 0.02 and 20 % respectively. The black step plot is the adaptation time distribution measured in experiments. (B) The adaptation time distribution for MBL (green) model extracted from the MaxEnt distribution when $\bar{\tau}$ and $\overline{\tau^2}$ are constrained to 245 s and 62323.5 s² respectively. The black step plot is the adaptation time distribution measured in experiments.

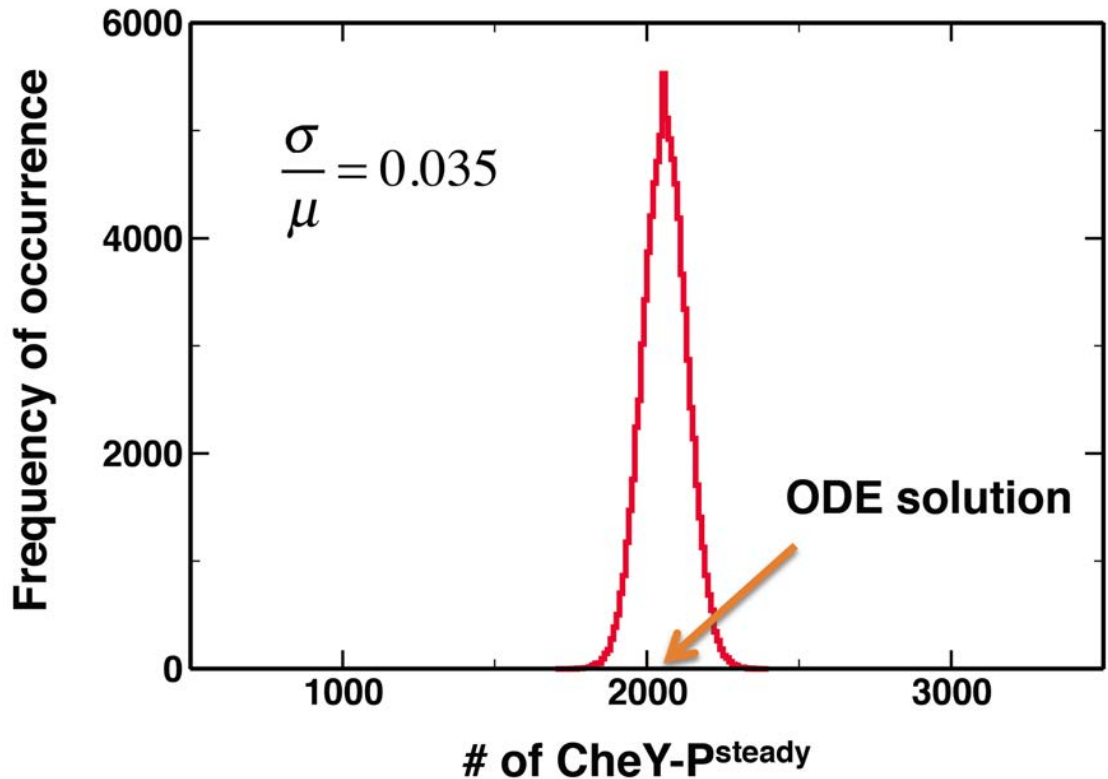

**Supplementary Figure 4: The role of intrinsic noise at pre-stimulus steady state:** For an in-silico cell with correlated under expression of all the chemotactic proteins to half of their average values, we have studied the role of intrinsic noise about the ODE solution for the steady state of CheY-P. Red curve shows the distribution of CheY-P values about the ODE solution indicated with the orange arrow. The noise is quantified as the ratio of the standard deviation to the mean of the distribution. We see that the standard deviation is about 30 times smaller than the mean.

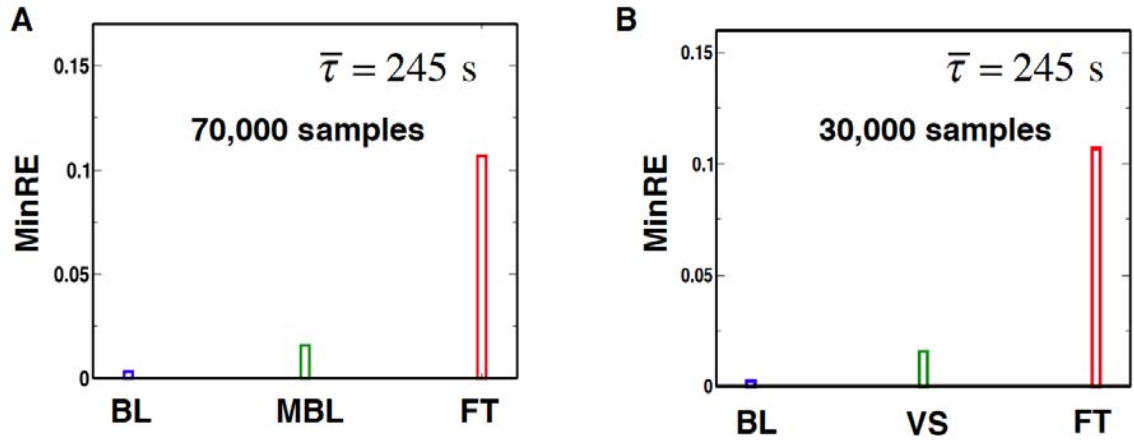

**Supplementary Figure 5: The convergence of MinRE: (A)** Bar plot of MinRE for the BL(blue), MBL (green) and the FT (red) when $\bar{\tau}$ = 245 s for a sample size of 70,000. **(B)** Bar plot of MinRE for the BL(blue), MBL (green) and the FT (red) when $\bar{\tau}$ = 245 s for a sample size of 30,000.

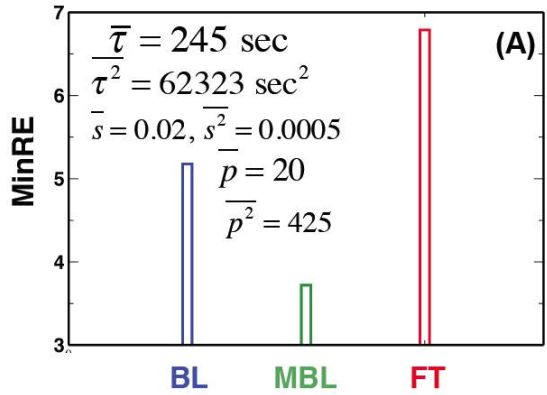

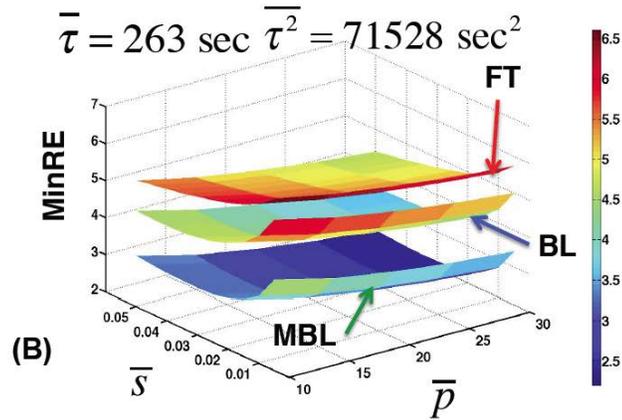

**Supplementary Figure 6:** **(A)** Shows MinRE values for the MBL, BL, and the FT model when the average values and the variances of the variables, $\tau$, $p$, and $s$ were constrained. The relative rank ordering of the models (MBL>BL>FT) based on the MinRE values do not change from the results shown in the main text. We used a variance of p that ensures that majority of the E. coli cells produce CheY-P abundances that are within the working range of the flagellar motors. The variance of s was not available from experiments and we used an arbitrary value for $\overline{s^2}$. **(B)** We studied the sensitivity of the rank ordering of the three models (FT, BL, and MBL) when the average values of $\tau$, $p$, and $s$ are changed in the constraint equations. Comparison of the results of this figure with that of Fig. 3E shows that changing the average values of $\tau$ and $\tau^2$ within the uncertainties of the estimated expected values for these variables does not change the relative rank ordering of the models. The errors in equating sample averages to expected values arise due to small sample sizes. These uncertainties can produce errors in the estimations of the Lagrange multipliers.

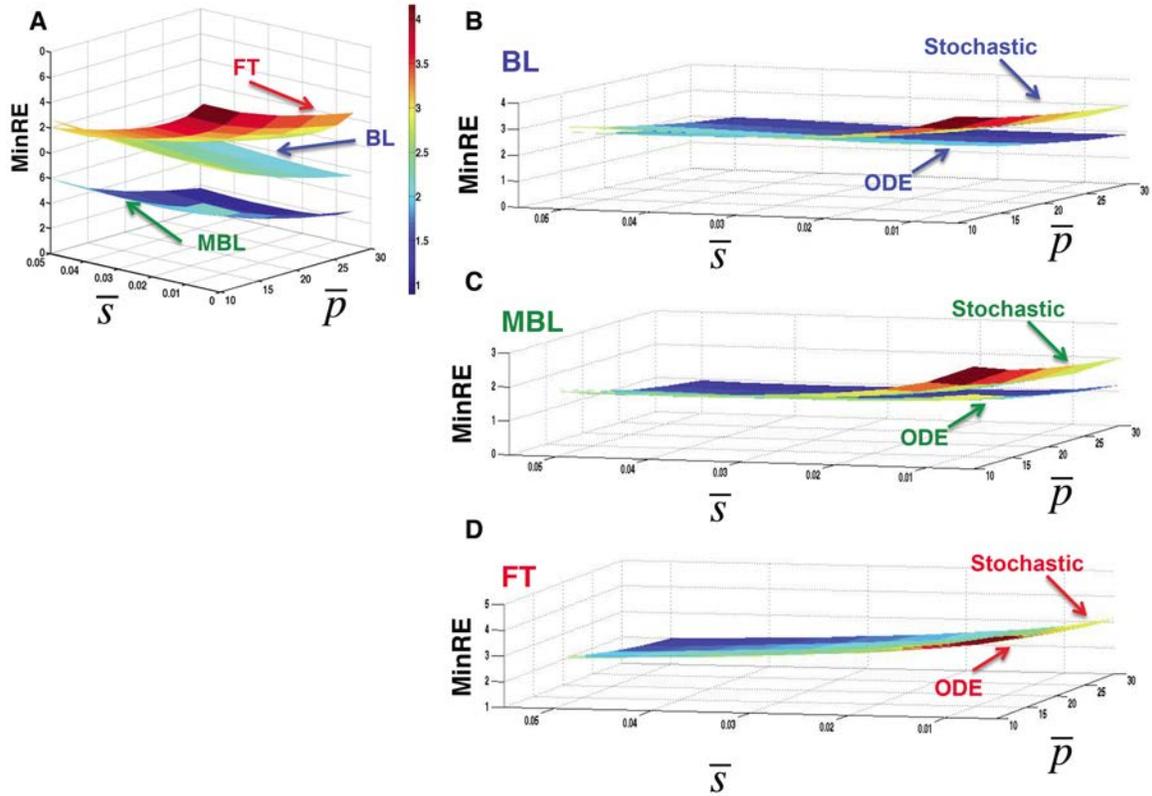

**Supplementary Figure 7: The role of intrinsic noise on MinRE: (A)** The MinRE is plotted for the BL, MBL and the FT models as a function of $\bar{s}$ and $\bar{p}$ for $\bar{\tau} = 245s$, when the intrinsic noise fluctuations are ignored and the trajectories are generated deterministically. **(B)** Comparison of the values of MinRE for the BL model when the trajectories were generated using stochastic simulations of the Master Equation to the case when the trajectories were generated using deterministic rate equations. The values of the constraints are the same as shown in **(A)**. **(C)** Same plots as **(B)** but for the MBL model. **(D)** Same plots as **(B)** but for the FT model.

# Derivation of Eq. 9 in the main text

We describe our method of inferring distributions of total abundances of protein species in individual E. coli cells below. We define the Shannon's entropy for the stochastic trajectories $\{\Gamma\}$ as,

$$S = -\sum_{\Gamma} P_\Gamma \ln P_\Gamma \tag{S1}$$

A stochastic trajectory, $\Gamma$, represents changes in the abundances of signaling proteins in an individual cell in a time interval $t_0$ to $t_n$ by a set ($\{n_j\}, t_n ; \{n_j\}, t_{n-1} ; \{n_j\}, t_{n-2} ;…;  \{n_j\}, t_1; \{n_j\}, t_0; \{n^{total}_q\}$) where copy numbers of different proteins, $\{n_j\}$ (j=1…$N_P$ = total # of distinct signaling proteins), are given the times, $t_{n-i} = t_0 + (n-i)\Delta$, i=0..n, where, $\Delta$ is smaller than or of the same order of the smallest reaction time scale (Fig. 2). q denotes the number of different protein species, $N_T$. $N_P \geq N_T$, as a protein species can be modified during signaling, e.g., the signaling protein CheY-P is generated from the protein CheY.

Therefore,

$$\begin{aligned}P_\Gamma &= P(\{n_j\},t_n; \{n_j\},t_{n-1}; \{n_j\},t_{n-2};\cdots|\{n_j\},t_0)P(\{n_j\},t_0; \{n_q^{total}\}) \\ &= P(\{n_j\},t_n; \{n_j\},t_{n-1}; \{n_j\},t_{n-2};\cdots|\{n_j\},t_0)P(\{n_j\},t_0|\{n_q^{total}\})P(\{n_q^{total}\})\end{aligned} \tag{S2}$$

$P(\{n_j\}, t_n ; \{n_j\}, t_{n-1} ; \{n_j\}, t_{n-2} ;…;  \{n_j\}, t_1 | \{n_j\}, t_0)$ is the conditional probability of producing the copy numbers of the signaling species in a stochastic trajectory $\Gamma$ at the times $\{t_n … t_1\}$, given there is a specific set of copy numbers of proteins ($\{n_j\}$) when attractants are added at the pre-stimulus state at $t_0$. $P(\{n_j\}, t_0; \{n^{total}_q\})$ denotes the joint probability of having the pre-stimulus state with specific copy numbers ($\{n_j\}$) at time $t_0$ with total protein concentrations $\{n^{total}_q\}$. This joint probability can be written as a product of the conditional probability $P(\{n_j\}, t_0|\{n^{total}_q\})$, describing the probability of having the specific pre-stimulus state at $t_0$ given a specific set of total protein abundances $\{n^{total}_q\}$, and the probability of occurrence of $\{n^{total}_q\}$ or $P(\{n^{total}_q\})$, i.e, $P(\{n_j\}, t_0; \{n^{total}_q\}) = P(\{n_j\}, t_0|\{n^{total}_q\}) P(\{n^{total}_q\})$.

The biochemical signaling reactions producing E. coli chemotaxis are described by Markov processes where the conditional probability, $P(\{n_j\}, t_p)\}| \{n_j\}, t_{p-1}\})$, for changing the signaling state of the system changes from $\{\{n_j\}, t_{p-1}\}$ to $\{\{n_j\}, t_p\}$ is given by the Master Equation(8),

$$\partial P(\{n_j\},t_p|\{n_j\},t_{p-1})/\partial t_p = L P(\{n_j\},t_p|\{n_j\},t_{p-1}) \tag{S3}$$

, where, $L$ describes a linear operator (8) dependent on the biochemical reaction rates, wiring of the signaling network, and the copy numbers of signaling proteins at time $t_{p-1}$. Therefore, the conditional probability, $P(\{n_j\}, t_n ; \{n_j\}, t_{n-1} ; \{n_j\}, t_{n-2} ;…;  \{n_j\}, t_1 | \{n_j\}, t_0) = P(\{n_j\}, t_n | \{n_j\}, t_{n-1})P(\{n_j\}, t_{n-1} | \{n_j\}, t_{n-2}) … P(\{n_j\}, t_1 | \{n_j\}, t_0)$ (equality holds for a Markov process), is entirely determined by the solutions of the above Master Equation

and the initial condition at t=t₀. We consider variations in $P_\Gamma$ arising from the variations in $P(\{n^{total}_q\})$, i.e.,

$$\delta P_\Gamma = P(\{n_j\},t_n;\{n_j\},t_{n-1};\{n_j\},t_{n-2};\cdots|\{n_j\},t_0)P(\{n_j\},t_0|\{n^{total}_q\})\delta P(\{n^{total}_q\}) = P_C\,\delta P(\{n^{total}_q\})$$

(S4)

where, we write $P_C = P(\{n_j\},t_n;\{n_j\},t_{n-1};\{n_j\},t_{n-2};\cdots|\{n_j\},t_0)P(\{n_j\},t_0|\{n^{total}_q\})$ to simplify notations.

We maximize S in Eq. (S1) when the total protein abundances are varied as above in the presence of the constraints given by Eq. (S5)-(S6). Below we show the derivation for two different constraints, which can be easily generalized to include additional constraints.

We consider cell population average of a chemotactic response produced by a stochastic trajectory (or an individual cell), $f_{a\Gamma}$, which depends on the entire trajectory, $\Gamma$, and, the average of total abundance of a particular protein, $n_1^{total}$. The above constraints are described in the equations below:

$$\overline{f_a} = \sum_\Gamma f_{a\Gamma} P_\Gamma \tag{S5}$$

$$\overline{n_1^{total}} = \sum_{\{n^{total}_q\}} n_1^{total} P(\{n^{total}_q\}) \tag{S6}$$

Maximization of the entropy, S, by $P_\Gamma = \hat{P}_\Gamma$ or $P(\{n^{total}_q\}) = \hat{P}(\{n^{total}_q\})$ will produce the equation below. For simplifying the notation we abbreviate, $P(\{n^{total}_q\})$ as $P_0$.

$$\delta S = 0 = \sum_\Gamma P_C(\ln P_C + \ln P_0)(\delta P_0) + \sum_\Gamma P_C(\delta P_0) \tag{S7}$$

Variations from the constraints equations will produce,

$$0 = \sum_\Gamma f_{a\Gamma} P_C(\delta P_0) \tag{S8}$$

and,

$$0 = \sum_{\{n^{total}_q\}} n_1^q(\delta P_0) \tag{S9}$$

Therefore,

$$\hat{P}(\{n^{total}_q\}) = Z^{-1}\exp(-\lambda_a \sum_{\Gamma_C} f_{a\Gamma} P_C - \eta_1 n_1^{total} - \sum_{\Gamma_C} P_C \ln P_C) \tag{S10}$$

where, $\{\Gamma_C\}$ describes a set of stochastic trajectories ($\{n_j\}$, $t_n$ ; $\{n_j\}$, $t_{n-1}$ ; $\{n_j\}$, $t_{n-2}$ ;....; $\{n_j\}$, $t_1$; $\{n_j\}$, $t_0$) with a fixed specific total protein abundances, $\{n_q^{total}\}$. Therefore, $\{\Gamma_C\}$ essentially denotes averages over variations of stochastic trajectories due to intrinsic noise fluctuations. $\lambda_a$ and $\eta_1$ are the Lagrange multipliers, which are determined by substituting Eq. (S10) in the equation for the constraints (Eqns S8 and S9) and $Z$ is the partition function. In deriving Eq. S10, we also used the fact, $\sum_{\Gamma_C} P_C = 1$.